\documentclass[fleqn,usenatbib]{mnras}

\usepackage{newtxtext,newtxmath}

\usepackage[T1]{fontenc}
\usepackage{ae,aecompl}

\newcommand{\avg}[1]{\langle#1\rangle}
 
\newcommand{\beq}{\begin{equation}}
\newcommand{\eeq}{\end{equation}}
\newcommand{\bit}{\begin{itemize}}
\newcommand{\eit}{\end{itemize}}

\newcommand{\beqa}{\begin{equation}\begin{aligned}}
\newcommand{\eeqa}{\end{aligned}\end{equation}}
\newcommand{\comment}[1]{}

\newcommand{\Msun}{\rm M_\odot}

\newcommand{\DS}{\Delta\Sigma}

\usepackage{graphicx}	
\usepackage{amsmath}	
\usepackage{array}      
\usepackage{subfigure}
\usepackage{soul}
\usepackage{booktabs} 
\usepackage{multirow}

\usepackage[dvipsnames]{xcolor}

\title{Incorporating galaxy cluster triaxiality in stacked cluster weak lensing analyses}
\author[Z. Zhang et al.]
{Zhuowen Zhang,$^{1,2}$\thanks{E-mail: zzhang13@uchicago.edu (UofC)}
Hao-Yi Wu,$^{3}$
Yuanyuan Zhang,$^{2}$ Joshua Frieman,$^{1,2}$
\newauthor
Chun-Hao To,$^{4}$ Joseph DeRose,$^{4,5,6}$ Matteo Costanzi,$^{7,8,9}$  Risa H. Wechsler,$^{4,5,6}$ 
\newauthor
Susmita Adhikari,$^{1}$
Eli Rykoff,$^{5,6}$ Tesla Jeltema,$^{10}$  August Evrard,$^{11}$ Eduardo Rozo$^{12}$ \\
$^{1}$Department of Astronomy and Astrophysics, University of Chicago, 5640 S. Ellis Ave, Chicago, IL, USA, 60637 \\
$^{2}$Fermilab, Kirk \& Pine Road, Batavia, IL, USA, 60510 \\
$^{3}$Department of Physics, Boise State University,  Boise, ID, USA, 83725\\
$^{4}$Department of Physics, Stanford University, 382 Via Pueblo Mall, Stanford, CA USA, 94305 \\ 
$^{5}$Kavli Institute for Particle Astrophysics $\&$ Cosmology, P. O. Box 2450, Stanford University, Stanford, CA, USA, 94305\\ 
$^{6}$SLAC National Accelerator Laboratory, Menlo Park, CA, USA, 94025 \\
$^{7}$Astronomy Unit, Department of Physics, University of Trieste, via Tiepolo 11, I-34131 Trieste, Italy \\
$^{8}$INAF-Osservatorio Astronomico di Trieste, via G. B. Tiepolo 11, I-34143 Trieste, Italy \\
$^{9}$Institute for Fundamental Physics of the Universe, Via Beirut 2, 34014 Trieste, Italy \\
$^{10}$Santa Cruz Institute for Particle Physics, Santa Cruz, CA, USA, 95064 \\
$^{11}$Department of Astronomy, University of Michigan, Ann Arbor, MI, USA, 48109 \\
$^{12}$Department of Physics, University of Arizona, Tucson, AZ, USA, 85721 \\
}

\date{Accepted XXX. Received YYY; in original form ZZZ}

\pubyear{2022}

\begin{document}
\label{firstpage}
\pagerange{\pageref{firstpage}--\pageref{lastpage}}
\maketitle

\begin{abstract}
Counts of galaxy clusters offer a high-precision probe of cosmology, but control of systematic errors will determine the accuracy, and thus the cosmological utility, of this measurement. Using Buzzard simulations, we quantify one such systematic, the triaxiality distribution of clusters identified with the redMaPPer optical cluster finding algorithm, which was used in the Dark Energy Survey Year-1 (DES Y1) cluster cosmology analysis. We test whether redMaPPer selection biases the clusters' shape and orientation and find that it only biases orientation, preferentially selecting clusters with their major axes oriented along the line of sight. We quantify the boosting of the observed redMaPPer richness for clusters oriented toward the line of sight. Modeling the richness--mass relation as log-linear with Poissonian intrinsic scatter, we find that the log-richness amplitude $\ln(A)$ is boosted from the lowest to highest orientation bin with a significance of $14\sigma$, while the orientation dependence of the richness-mass slope and intrinsic scatter is minimal. We also find that the weak lensing shear-profile ratios of cluster-associated dark halos in different orientation bins resemble a ``bottleneck'' shape that can be quantified with a Cauchy function. We test the correlation of orientation with two other leading systematics in cluster cosmology---miscentering and projection---and find a null correlation, indicating that triaxiality bias can be forward-modeled as an independent systematic. Analytic templates for the triaxiality bias of observed-richness and lensing profiles are mapped as corrections to the observable of richness-binned lensing profiles for redMaPPer clusters. The resulting mass bias confirms the DES Y1 finding that triaxiality is a leading source of bias in cluster cosmology. However, the richness-dependence of the bias confirms that triaxiality, along with other known systematics, does not fully resolve the tension at low-richness between DES Y1 cluster cosmology and other probes. Our model can be used for quantifying the impact of triaxiality bias on cosmological constraints for upcoming weak lensing surveys of galaxy clusters.
\end{abstract}

\begin{keywords}
cosmology: theory ---
cosmological parameters ---
galaxies: clusters: general ---
gravitational lensing: weak
\end{keywords}

\newpage
\section{Introduction}
The growth of the most massive structures in the universe is a sensitive probe of the $\mathrm{\Lambda CDM}$ cosmological model. Within this model, the number of dark matter halos of a given mass, or the halo mass function, depends sensitively  upon the current matter density, $\Omega_m$, and on the linear density fluctuation amplitude at the 8 $h^{-1}$~Mpc scale, $\sigma_8$. Beyond $\Lambda$CDM, the halo mass function is also sensitive to the dark energy equation of state parameter, $w$ \citep[see, e.g.,][for reviews]{Frieman08, Weinberg13, Huterer15}.  

Comprising a few to hundreds of galaxies, galaxy clusters are tracers of and proxies for dark halos in the approximate mass range  $10^{13} -  3\times10^{15}~h^{-1}$~M$_{\odot}$. Since the mass of a galaxy cluster is difficult to directly observe, it is typically inferred from another cluster observable through a mass-observable relation (MOR).  
Examples of such observables are the number counts of galaxies per cluster, often referred to as the ``richness" \citep{Koester07, Rykoff14}; X-ray emission luminosity or temperature from the intracluster medium \citep[ICM;][]{Piffaretti11, Mehrtens12}; and the inverse Compton scatter parameter of Cosmic Microwave Background photons off of the ICM electrons, known as the Sunyaev-Zel'dovich effect \citep{Planck16, Bleem15}. The precision of cluster cosmology studies relies on an accurate statistical model relating these observables to cluster mass \citep{Allen11}. 

The Dark Energy Survey (DES)  used the 4-m Blanco Telescope and the Dark Energy Camera \citep{Flaugher15} to carry out a multi-band, 5,000 deg$^2$ survey over six years, with the primary goal of constraining cosmology and the nature of dark energy. Given its depth and wide-area coverage, DES observed $\sim$100,000 galaxy clusters up to redshift $\sim 1$ \citep{Melchior17}. Initial cluster cosmology results, based on the first year of data (DES Y1), were published in \cite{DESY1_Cluster}.
The cluster observable that DES Y1 employed as a mass proxy is a probabilistic cluster galaxy count called richness, computed with the redMaPPer algorithm \citep{Rykoff12}. 

Gravitational lensing, the shearing of galaxy images by foreground mass concentrations, is one of the most powerful methods for calibrating cluster mass-observable relations  \citep{Johnston07, Gruen14, Simet16, McClintock18}. 
DES calibrates the cluster MOR through statistical weak lensing, in which shears from an ensemble of clusters are stacked to achieve high signal-to-noise \citep{Bartelmann01}. In DES, stacked shear profiles are estimated for clusters binned in redMaPPer richness, enabling a determination of the mean halo mass as a function of richness \citep{Melchior17, McClintock18}. 

Systematic effects in cluster selection or in calibration of the cluster MOR, if uncorrected for, can lead to biased cosmological inference from cluster abundance measurements. One such systematic arises from cluster triaxiality, the intrinsically elliptical shapes of galaxy clusters. N-body simulations indicate that dark halos can have major-to-minor axis ratios as high as 1.5 \citep{Jing&Suto02, Oguri05}, as confirmed observationally through cluster weak lensing ellipticity measurements \citep{Clampitt16, Shin17}. Failing to account for cluster halo triaxiality may result in an overestimate of cluster mass by as much as 3-6\% for stacked weak lensing measurements \citep{Dietrich14}. Triaxiality was identified as one of the most important sources of systematic bias in the DES Y1 cluster lensing analysis, significant at the 2\% level
\citep{McClintock18}. Recently \cite{Osato18} showed that triaxiality not only biases the cluster surface mass density in the ``one-halo" regime but also affects the surface density profile in the ``two-halo" regime. 

In this paper, we use redMaPPer cluster samples and associated halo catalogs in the Buzzard simulations to quantify cluster selection bias related to halo triaxiality properties such as orientation and ellipticity. We evaluate the impact of the triaxiality selection bias on 1) the richness--mass relation and 2) the excess surface mass density of individual halos \citep{Osato18}. The stacked surface density profiles modeled with a triaxiality selection bias deviate from the isotropically stacked profiles;   we find results comparable to those previously reported in the literature. 

The paper is organized as follows. In Section 2, we describe the simulation data set used in the study and the halo--cluster matching algorithm. In Section 3 we examine the orientation and ellipticity distributions of triaxial halos associated with redMaPPer-selected clusters, quantifying the preference for halo orientation along the line of sight. In Section 4 we examine the boost in cluster richness for a given mass resulting from this orientation selection bias in the cluster sample. In Section 5 we test for correlation of halo triaxiality with other leading systematics, finding no evidence for such. In Section 6 we study halo surface mass densities as a function of orientation and the effect of orientation selection bias on stacked surface density measurements. We conclude in Section 7.  

Throughout, we assume a flat $\Lambda$CDM cosmology with $\Omega_m = 0.283$, and $H_0 = 70$~km~s$^{-1}$~Mpc$^{-1}$. Distances and masses, unless otherwise noted, are defined in units of $h^{-1}$~Mpc and $h^{-1}$~M$_{\odot}$. 

\section{The Simulation Data set}
\subsection{Buzzard simulations}
We make use of the DES N-body simulation catalogs from the suite of Buzzard simulations \citep{DeRose19} with the $\Lambda$CDM parameters given above. Detailed descriptions of the simulations can be found in  \cite{MacCrann18, DeRose19, Wechsler21}; here we present a brief overview. 

The Buzzard simulations simultaneously achieve good spatial resolution and large volume by dividing the lightcone into three simulation boxes covering the redshift ranges $z \in [0.0, 0.34)$, $[0.34, 0.90)$, and $[0.90, 2.35)$, with respective minimally resolved dark matter particle masses of $2.7\times10^{10}~h^{-1}$~M$_{\odot}$,  $1.3\times10^{11}~h^{-1}$~M$_{\odot}$, and $4.8\times10^{11}~h^{-1}$~M$_{\odot}$. The increased resolution at low redshift captures non-linear structures at late times, while the lower resolution at high redshift enables the catalogs to encompass larger total volume. Particles are evolved using the L-Gadget2 code designed to efficiently run large-volume dark-matter only simulations \citep{Springel05}.

Halos are found by ROCKSTAR  \citep{Behroozi13} with masses defined by $M_{200b}$, the mass enclosed in a radius within which the average matter density is 200 times the mean matter density of the universe at the halo redshift. Galaxies are assigned to dark matter particles using ADDGALS, an empirical algorithm that places galaxies on dark matter particles based on a galaxy--dark matter relation learned from subhalo abundance matching catalogs and that is designed to accurately reproduce galaxy luminosities, colors, and spatial clustering over large volumes \citep{DeRose19}. In particular, each massive halo is probabilistically assigned a luminous, red galaxy at its center. 

\subsection{redMaPPer cluster sample}
With the advent of wide-field-imaging surveys, a plethora of optical cluster finding algorithms have emerged, such as those based on galaxy photometric redshifts, e.g. \cite{Kepner00}, \cite{Soares-Santos11}, \cite{Wen12}, and  \cite{Oguri14}. In this paper, we study the cluster sample identified with the redMaPPer algorithm \citep{Rykoff14}, which identifies cluster candidates as spatial over-densities of red-sequence galaxies. Clusters are assumed to be centered on a galaxy, with the central galaxy selected based on its luminosity and color (brightest central galaxy, or BCG). The algorithm also produces a richness estimate, $\lambda$, for each cluster candidate, a probabilistic count of cluster red-sequence galaxies above a luminosity threshold and inside a spatial aperture determined from iterative richness estimations. It uses a sample of observed clusters with spectroscopic redshifts as a training set to build the initial redshift-dependent red-sequence model which cluster galaxies are fitted onto to determine the photometric redshift $z_{\lambda}$.

The redMaPPer cluster finder has been applied to the Buzzard catalogs to identify galaxy clusters. We make use of a redMaPPer sample with a richness cut $\lambda > 20$ to ensure the purity of the sample, and a maximum cluster redshift of $z<0.90$ \citep{Rykoff16, McClintock18} which is around the redshift detection limit of redMaPPer and the limit of the Buzzard light cone. Halos are also cut at masses below $5 \times 10^{13}~h^{-1} M_{\odot}$ which roughly corresponds to a richness of 20.

\subsection{Cluster halo matching algorithm}
\label{sec:cl_sel} 

Here we outline how redMaPPer clusters are matched to Buzzard halos. First, 
a cluster is labeled as centered or miscentered based on whether or not its redMaPPer BCG is a central galaxy in a Buzzard halo. Centered clusters have BCGs that share the same ID as that of the halo central galaxy; in this case, the cluster and halo central coordinates perfectly match.  
By this criterion, 63\% of redMaPPer clusters are centered; the remaining were matched using the halo-cluster algorithm described below. A more detailed description of the centering properties of the redMaPPer catalogs can be found in Section \ref{subsec:miscentering}.

The miscentered redMaPPer clusters were matched to Buzzard dark matter halos by proximity. Halos were ranked by halo mass, and clusters were ranked by richness, both in descending order. We first search for halo-cluster pairs with redshift separation $\Delta z \leq 0.05$ between cluster photometric redshift and true halo redshift. This range of redshift separation is large compared to the typical photometric redshift error, $\Delta z \sim  0.005$, for redMaPPer-selected clusters. Then, for each halo, we identify those redMaPPer clusters with BCGs within a projected 2-D, comoving radius of $2~h^{-1}~\mathrm{Mpc}$ of the halo central galaxy. If there are multiple redMaPPer clusters satisfying these separation criteria, we match the halo to the richest such cluster that hasn't been previously matched. For each cluster, we repeat this matching process, selecting halos satisfying the redshift and projected distance criteria, and then choosing the most massive such halo still on the list as the one to be associated with that cluster. Clusters and halos that uniquely match with each other in both matching steps are considered valid matches. 

Of the 24,243 initially identified redMaPPer cluster candidates in the suite of 18 catalogs, 23,658 or 97\% are uniquely matched to a halo with the above prescription. We do not consider the 
non-uniquely matched clusters in this study.

This halo--cluster matching algorithm was cross-checked with an independent halo--cluster matching algorithm used in \cite{Farahi16} that rank-orders halos and clusters by the number of galaxies they have in common. Using the Aardvark simulation, \cite{Farahi16} uniquely matched 99\% of redMaPPer clusters to halos, showing excellent agreement with this paper's algorithm on the completeness and uniqueness of cluster-to-halo matches. We cross checked our matching algorithm with that of \cite{Farahi16} in a different version of Buzzard with a smaller patch of sky containing several hundred clusters and found almost identical halo-cluster pairings.

Due to the high number of particles per halo, Poisson noise plays a negligible role in our ellipticity measurements: at low redshift, with a mass resolution of $2.7 \times 10^{10}$~$h^{-1}$~M$_\odot$, a typical $3 \times 10^{14}$~$h^{-1}$~M$_\odot$-mass halo found through redMaPPer corresponding to a richness of $\sim 40$ will contain $\sim 10,000$ particles, and the same-mass halo at high redshift, with a poorer mass resolution of $1.3 \times 10^{11}~$~$h^{-1}$~M$_\odot$, contains $\sim 3,000$ particles. Simulations conducted by \cite{Jing&Suto02} demonstrated that these large numbers of particles per halo make Poisson noise negligible for our purposes. We do not consider halos with fewer than $100$ particles with poor shape convergence, corresponding to group size objects with richnesses well below our $\lambda > 20$ cut.

\begin{figure}
    \centering
    \subfigure{\includegraphics[width=0.45\textwidth]{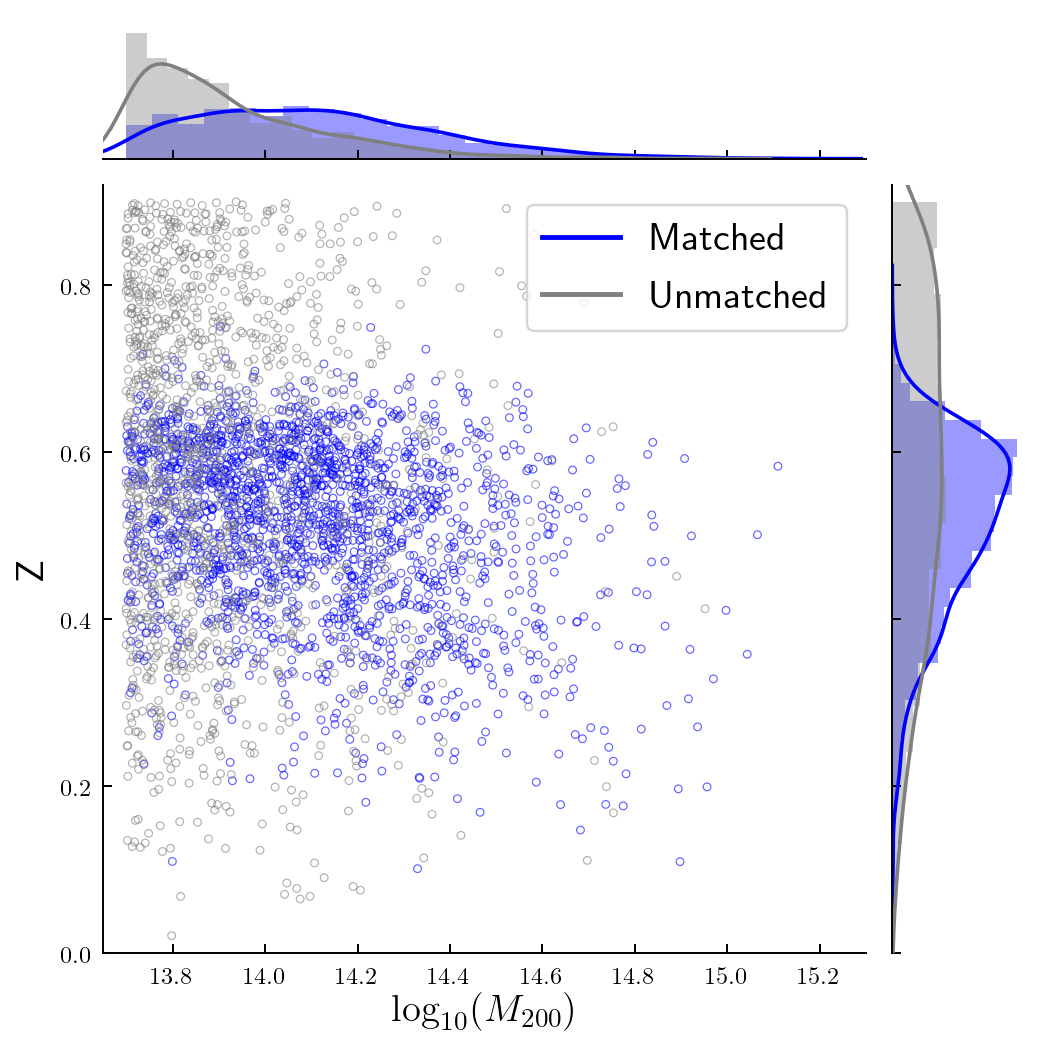}}
    \\
    \subfigure{\includegraphics[width=0.45\textwidth]{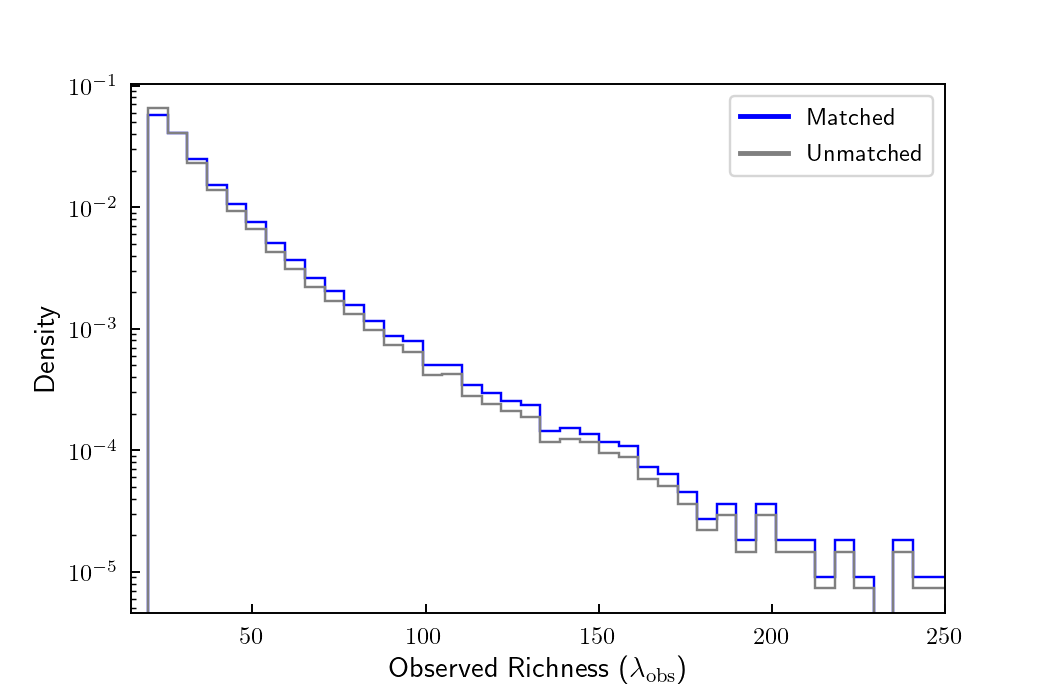}}
    \caption{{\it Upper panel}: A 2-D distribution plot of the true $M_{200m}$ and $z$ of halos before and after matching with redMaPPer clusters. The halos are cut at $M_{200m} > 5\times10^{13}~h^{-1}M_{\odot}$ and a redshift cut of $z < 0.90$ and are sparsely sampled for better visualization. {\it Lower panel}: The probability density function of the observed richness $\lambda_{\rm obs}$ before and after matching with halos. Because of the high match rate of redMaPPer clusters the two distributions are nearly identical. }
    \label{fig:M_z_lambda_dist}
\end{figure}

\section{Cluster Halo Triaxiality and Selection Bias
}
\label{sec:shape_orientation}
Previous studies have shown that optical cluster finders  preferentially select halos with their major axes oriented along the line of sight \citep{Corless08, Dietrich14}. In this Section, we  quantify this orientation bias of selected clusters using the redMaPPer catalogs and the Buzzard simulations. 
We also explore whether a cluster ellipticity selection effect exists, i.e., whether redMaPPer preferentially selects 
halos that are more or less elliptical than randomly selected halos.

\subsection{Measurement of Halo Ellipticity and Orientation}
We make use of a quadrupole moment tensor method (\cite{Bett12} and references therein) to measure the shapes and orientations of halos. Many such algorithms solve for halo shapes by using particles inside a spherical envelope \citep{Dietrich14, Osato18}; this has the advantage of allowing easy comparison with other results, but it systematically underestimates the axial ratios for ellipsoidal profiles, an effect known as ``edge bias.'' As described below, we correct for such an effect by using an iterative method to determine the shape of the enclosing envelope, in the vein of earlier works such as \cite{Dubinski91}, \cite{Katz91} and \cite{Warren92}. To do so, we first measure the shape of the halo using particles inside a spherical envelope; once the axis ratios and the principal axes are found, the envelope adapts iteratively until both the axis ratios of the halo inside the envelope and the shape of the ellipsoidal envelope itself converge. 

We now describe the halo ellipticity measurement algorithm in detail. It involves nested iteration of both the principal axes, as determined from the quadrupole moment tensor, and of the envelope shape. In the initial iteration, $l=0$, of the envelope shape, the envelope is set to be a sphere centered on the halo center with a radius equal to the virial radius of the halo, $R_{\rm vir}$. The reduced quadrupole moment tensor is then calculated for the $N_P$ dark matter particles inside the envelope. This tensor, with its principal-axis directions solved at the $k$-th iteration, is defined as:
\begin{equation}
\mathcal{M}^{(k)}_{ij} = \frac{1}{N_P^{(k)}}\sum_{p=1}^{N_p^{(k)}} \frac{R_{p,i}^{(k)} R_{p,j}^{(k)}}{\big(R_p^{(k)}\big)^2},
\end{equation}
where $R_{p,i}$ and $R_{p,j}$ are the distances from the center along Cartesian coordinate axes of the $p$-th particle and $R_p^{k}$ is the triaxial radius, defined below, of the $p$-th particle solved at the k-th iteration.

We define $a$, $b$, and $c$ as the major, intermediate, and minor axes lengths of a particle projected onto the unit sphere and $q \equiv \frac{c}{a}$ and $s \equiv \frac{b}{a}$ as the minor-major and intermediate-major axis ratios; the physical distances to the $p$-th particle along the minor, intermediate and major axes are denoted $X_p$, $Y_p$ and $Z_p$. In this notation, the triaxial radius at the k-th iteration of the particle is expressed as:
\begin{equation}
R_p^{(k)} = \sqrt{\Bigg(\frac{X_p}{q^{(k-1)}}\Bigg)^2 + \Bigg(\frac{Y_p}{s^{(k-1)}}\Bigg)^2 + Z_p^2}.
\end{equation}
The axis lengths projected onto the unit sphere are the square roots of the eigenvalues of the reduced tensor, and the axis directions are the corresponding eigenvectors. After each iteration, the principle axes are rotated by the rotation matrix $M^{(k)}$, where each row in the matrix is a principle axis found from the reduced tensor in the previous iteration. The reduced tensor is computed again under the rotated coordinates. Starting from $q^{(k=0)}=1$ and $s^{(k=0)}=1$, the tensor is considered to have converged if 
\begin{equation}
\left| 1-\frac{q^{(k)}}{q^{(k-1)}} \right| < 10^{-6}\quad \text{and} \quad \left| 1-\frac{s^{(k)}}{s^{(k-1)}}\right| < 10^{-6} ~~,
\label{eqn:ell_converge}
\end{equation}
and is deemed divergent if convergence is not reached before the number of iterations $k$ exceeds 100.

The total rotation matrix after $n$ rotations is 
\begin{equation}
\boldmath{M_{\rm tot}} = \boldmath{M^{(n)}}\ldots\boldmath{M^{(k)}}\ldots\boldmath{M^{(1)}}~,
\end{equation}
where each row in $M_{\rm tot}$ gives the direction of the corresponding halo axis prior to rotation.

If after $k$ iterations the axis ratios derived from the tensor converge, then the elliptical envelope of the particles is advanced from the previous $l-1$-th to the $l$-th (for $l>0$) iteration, adapting its axis ratios and orientation to those of the halo as determined from the tensor with the previous envelope. Particles with elliptical distances of 
\begin{equation}
R_p^{(l)} \equiv \sqrt{\Bigg(\frac{X_p^{(l-1)}}{q^{(l-1)}}\Bigg)^2 + \Bigg(\frac{Y_p^{(l-1)}}{s^{(l-1)}}\Bigg)^2 + (Z_p^{(l-1)})^2} < R_{\rm vir} 
\end{equation}
are selected. The sequence initializes at $q^{(l=0)} = s^{(l=0)} = 1$, and ($X_p^{0},Y_p^{0},Z_p^{0}$) along the original $(x,y,z)$ axes of our coordinate system and converges using the same criteria as for the shape of the halo inside the envelope, Cf. equation $\ref{eqn:ell_converge}$. The shape of the halo is said to be convergent only if both the shape of the halo particles found inside the envelope and the shape of the envelope itself both converge. 

\begin{figure}
	\centering
	\includegraphics[width=0.4\textwidth]{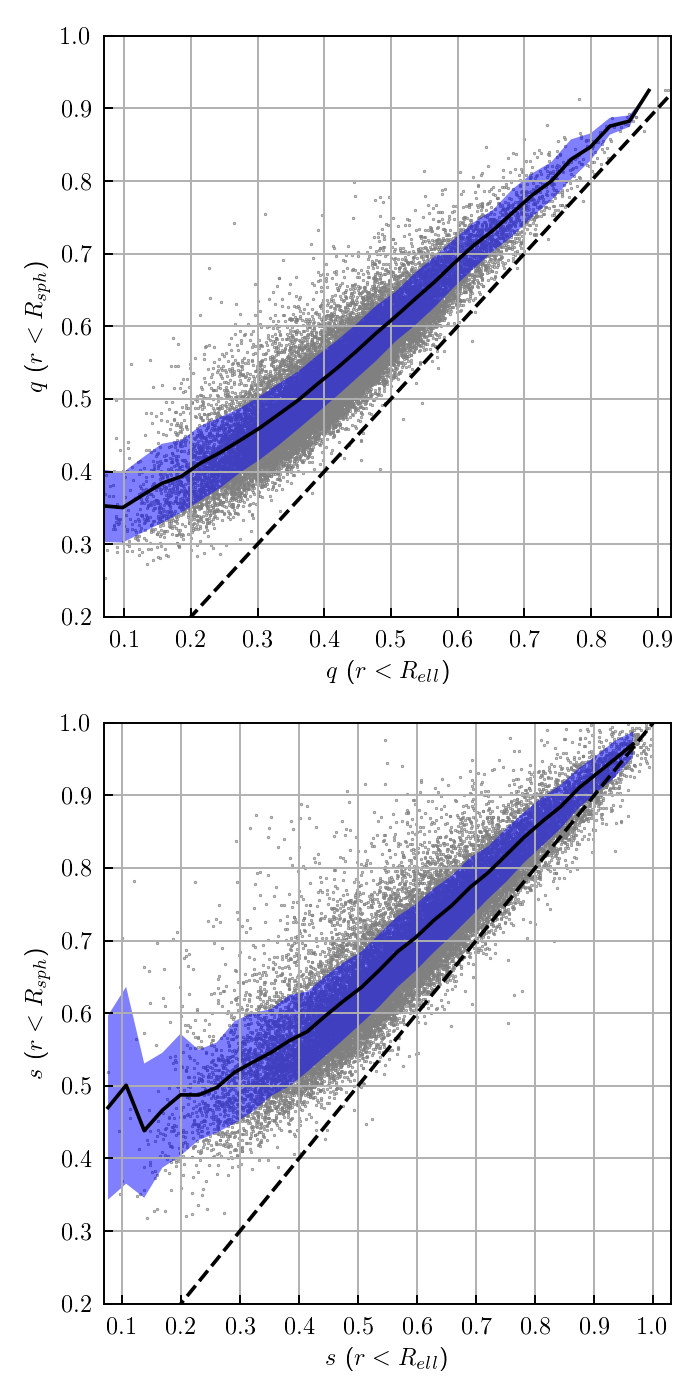}
    \caption{Axis ratios, $q$ and $s$, for redMaPPer-matched halos measured with spherical vs. adaptive ellipsoidal envelopes. Solid black lines show the mean ratios in each axis-ratio bin, and the blue bands indicate the $1-\sigma$ scatter. Dashed lines would correspond to no difference in axis ratios between the two methods. The results demonstrate 
    that edge bias reduces the measured ellipticities of halos from their true values, with larger bias at higher ellipticities 
    (smaller $q$ and $s$).  }
    \label{fig:edge_bias}
\end{figure}

We applied this technique to measure the shapes of simulated halos that are matched to the redMaPPer clusters; of the 23,658 matched redMaPPer clusters, the halo shape measurements converge by the above criteria for 22,790 of them. We use this sample in the following sections to explore orientation bias.

We can gauge the impact of the edge bias on halo shape measurement by comparing results with the adaptive ellipsoidal envelope to those using a fixed spherical envelope. In Fig. \ref{fig:edge_bias}, we plot the halo axis ratios $q$ and $s$ found using spherical envelopes (ordinates) with those from the adaptive ellipsoidal envelopes (abscissas). We see clearly that the axis ratios are biased high (ellipticities biased low) when using spherical envelopes, with larger bias at higher ellipticities (lower values of the axis ratios). These results are in qualitative agreement with those of \cite{Shin17}, who studied 2-D projected ellipticities of observed galaxies in redMaPPer clusters. They found that the inferred 2-D ellipticity, $e \equiv (1+q)/(1-q)$ where $q$ is the axis ratio for a 2-D ellipse, deviates by as much as 0.1 when using a circular aperture for the redMaPPer \citep{Rykoff14} cluster finder, 
$R_{\lambda} = 1 \text{$h^{-1}$~Mpc}(\lambda/100)^{0.2}$,
due to the cut-off of satellite galaxies along the major axis; they also found  that the bias in ellipticity becomes worse at higher ellipticity (smaller $q$). 

\subsection{Distributions of cluster halo orientation and ellipticity}
\label{subsec:bias_ellip_orienation}
\begin{figure*}
	\centering
	\includegraphics[width=0.8\textwidth]{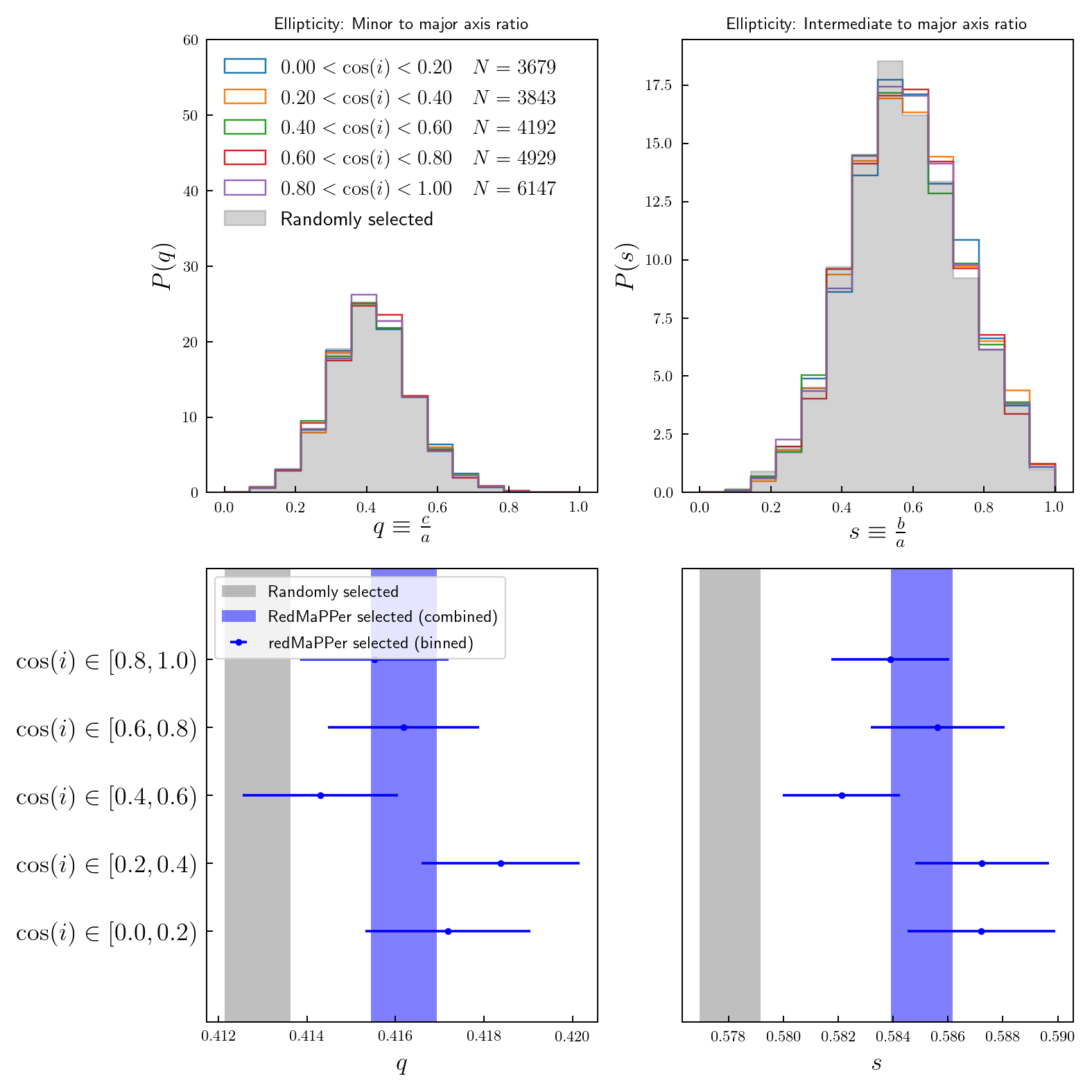}
    \caption{\textit{Top panels}: Axis-ratio distributions for redMaPPer-matched clusters binned by orientation and for randomly selected halos from the Buzzard simulations. \textit{Bottom panels}: Mean axis ratios with $1\sigma$ errors from jackknife resampling. Applying the $3\sigma$ significance cutoff rule, no significant shift is found in the shape parameters $q$ and $s$ for redMaPPer-matched and randomly selected halos. Also is the case that no statistically significant difference is found in the mean ellipticities across different orientation bins. }
    \label{fig:ellipticity_cosi}
\end{figure*}

Armed with measurements of halo shapes for redMaPPer clusters, 
in this subsection we study the distributions of halo ellipticity and orientation. To test for redMaPPer-associated selection biases, we compare these distributions to those for a sample of 36,445 randomly selected halos with convergent shape measurements from the Buzzard catalog. 
The orientation of interest is the angle between the halo major axis and the line of sight, which we denote by $i$; a non-uniform distribution of $i$ would signal the preferential selection of (prolate) clusters with these vectors aligned. For this analysis, we adopt the orientation bins $\cos(i) \in [0.0, 0.2),~[0.2, 0.4),~[0.4, 0.6),~[0.6,0.8),~[0.8, 1.0)$.

The distributions of axis ratios for redMaPPer-matched halos and for randomly selected halos are shown for different orientation bins in the upper panels of Figure \ref{fig:ellipticity_cosi}. Previous N-body studies found that more massive halos tend to be more elliptical \citep{Kasun05} as a result of tidal forces and mergers. To account for this effect, we resampled the randomly selected halos to match the halo mass function of the redMaPPer-matched halos. The upper panels of Fig. \ref{fig:ellipticity_cosi} indicate that the ellipticity distributions of the redMaPPer-matched halos are qualitatively very similar to those for the resampled random halos, with little dependence on orientation.

To quantify this comparison, in the bottom panels of Fig. \ref{fig:ellipticity_cosi} we show the mean axis-ratios for the redMaPPer-matched halos in different orientation bins (in blue), along with the means for the random halos (in grey). The errors on these measurements are estimated by jackknife resampling, with the simulated survey footprint split by the k-means algorithm  \textit{kmeans\_radec}\footnote{Code written by Erin Sheldon. Source: \url{https://github.com/esheldon/kmeans_radec}} into 40 non-overlapping patches---the error estimates come from the variance among the patches, each of them $37.5$ square degrees. With this kind of spatial jackknife, the choice of the size of the jackknife patch is a compromise: for very large patch size, the number of patches (samples) would be too small to get a meaningful statistical sample; for very small patch size, large-scale structure would be highly correlated across adjacent patches, so they could not be treated as quasi-independent for error estimation.

The mean axis ratios differ by 0.7 and 1.2\% for $q$ and $s$ respectively for redMaPPer vs. random halos. To determine if these differences are significant, we conduct a null-hypothesis test on $q$ and $s$ with their standard errors modeled as Student's $t$ distributions. We find a $1.4\sigma$ difference in the minor-to-major axis ratio $q$ for redMaPPer vs. randomly sampled halos and a $1.8\sigma$ difference in the intermediate-to-major axis ratio $s$. There are no statistically significant shifts in mean axis ratios for redMaPPer halos between different $\cos(i)$ bins. Thus, we do not find strong evidence of shifts in the ellipticity distributions. 

\begin{figure}
	\centering
	\includegraphics[width=0.5\textwidth]{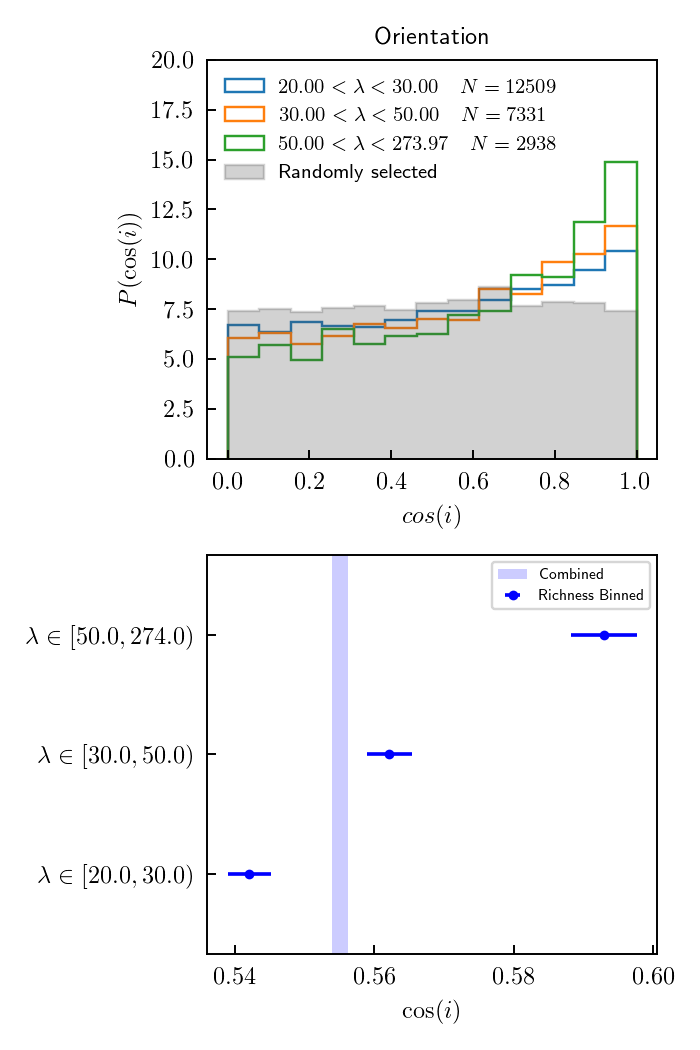}
    \caption{\textit{Top panel}: Distribution of $\cos(i)$ 
    for redMaPPer-matched halos in 3 richness bins and for randomly selected halos. \textit{Bottom panel}: The mean $\cos(i)$ for redMaPPer-selected halos is boosted relative to that for randomly selected halos (0.50, not shown).  The mean value of $\cos(i)$ also increases with redMaPPer richness. 
    Errors are estimated from jackknife resampling.}
    \label{fig:orientation}
\end{figure}

Figure \ref{fig:orientation} (top panel) shows a similar analysis to that above, but now for the distribution of halo orientation in 3 different richness bins. In this case, there is a clear signal of orientation bias in the redMaPPer-matched clusters, with preferential selection of clusters with major axis oriented along the line of sight. The effect is more pronounced for clusters of higher richness: the lower panel shows an increase in the mean value of $\cos(i)$ with richness. 
Using the same method of null hypothesis testing, we find that the mean value of $\cos(i)$ for redMaPPer halos of $0.555 \pm 0.002$ is boosted compared to that for randomly selected halos with a $13.8\sigma$ significance. There is also a statistically significant shift in the mean value of $\cos(i)$ between richness bins: the mean $\cos(i)$ for $\lambda \in [30.0, 50.0)$ ($\lambda \in [50.0, 274.0)$) exceeds that for $\lambda \in [20.0, 30.0)$ at $3.7\sigma$ ($4.8\sigma$) significance.  As a null test, we find that the randomly selected halos have a mean $\cos(i)$ consistent with 0.50.

In the next subsection,  we will interpret the correlation of mean $\cos(i)$ with richness seen in Fig. \ref{fig:orientation} as due to the boosting of observed richness for clusters (of fixed mass) oriented along the line of sight.

\section{Effect of orientation on the richness--mass relation}
\label{sec:richness--mass}

\begin{figure*}
    \centering
    \subfigure{\includegraphics[width=0.80\textwidth]{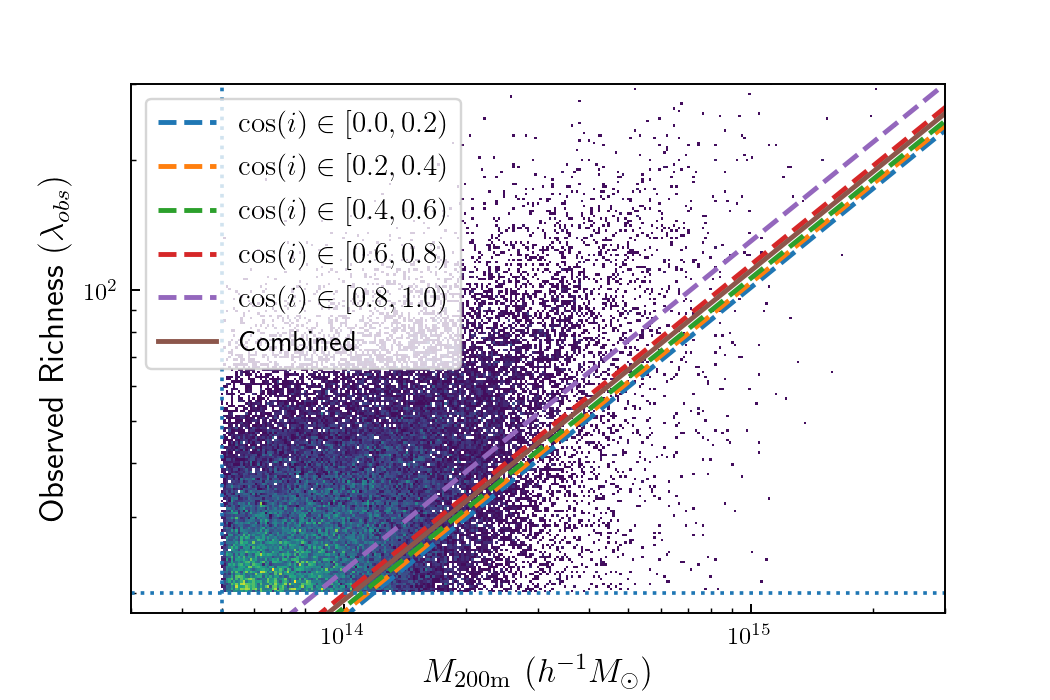}}
    \subfigure{\includegraphics[width=0.18\textwidth]{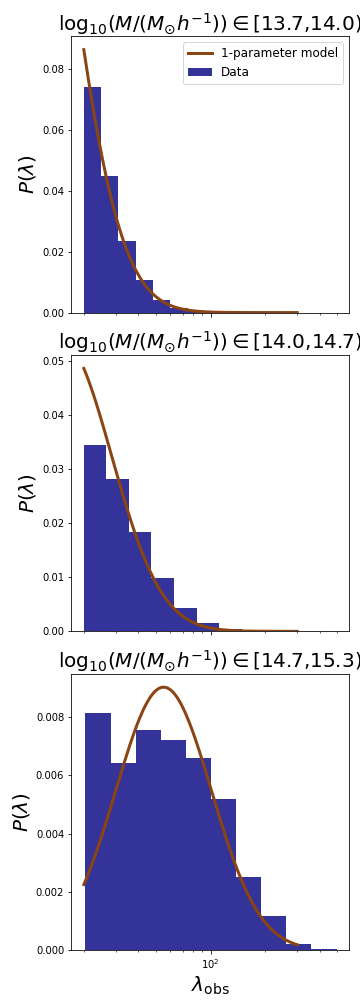}}
    \caption{\textit{Left panel:} Solid line labelled "Combined" shows the best-fit model to the full sample assuming a linear relationship between $\ln(\lambda)$ and $\ln(M)$. Dashed lines show best-fit models in each orientation bin, with the amplitude $\ln(A)$ allowed to vary from bin to bin. For halos of fixed mass, those oriented along the line of sight have larger observed redMaPPer richness. The dashed horizontal line indicates the richness cut at $\lambda >20$ and dashed vertical line the mass cut at $M > 5\times 10^{13}~h^{-1}M_{\odot}$. Color coded is the density of the scatter points in the parameter space, with brighter colors indicating a higher density. \textit{Right panel:} The richness distribution in mass bins for all data points overlaid with a truncated Gaussian fit using the best-fit parameters in the "Combined" 1-parameter model. In lower mass bins the best fit mean log-richness $\mu(\ln{\lambda})$ is lower than the mean log-richness of the data points, as the peak of the truncated Gaussian fit lies below the $\lambda > 20$ cutoff.}
    \label{fig:mass_richness_cosi}
\end{figure*}

\begin{figure}
	\includegraphics[width=0.5\textwidth, ]{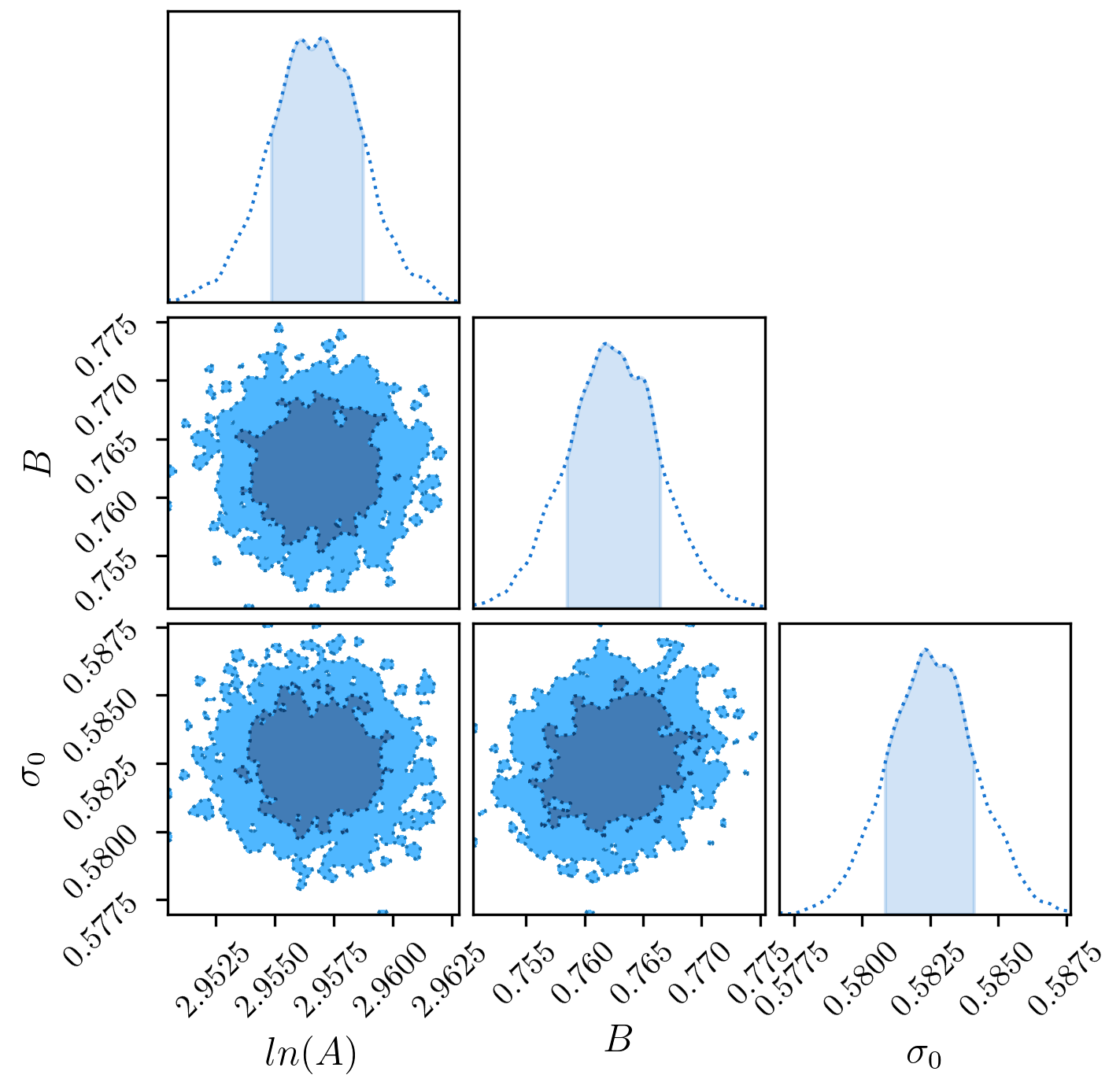}		
    \caption{Posterior distributions of the richness--mass parameters derived using all redMaPPer-matched clusters. The shaded regions in the 2-D distributions show the 68 and 95\% confidence regions; shaded regions in 1-D plots indicate the 68\% confidence regions for the marginalized parameters. Posteriors for templates in different orientation bins share the same features.
    }
    \label{fig:cornerplot_richness_mass}
 \end{figure}

\begin{figure}
	\centering
	\includegraphics[width=0.4\textwidth]{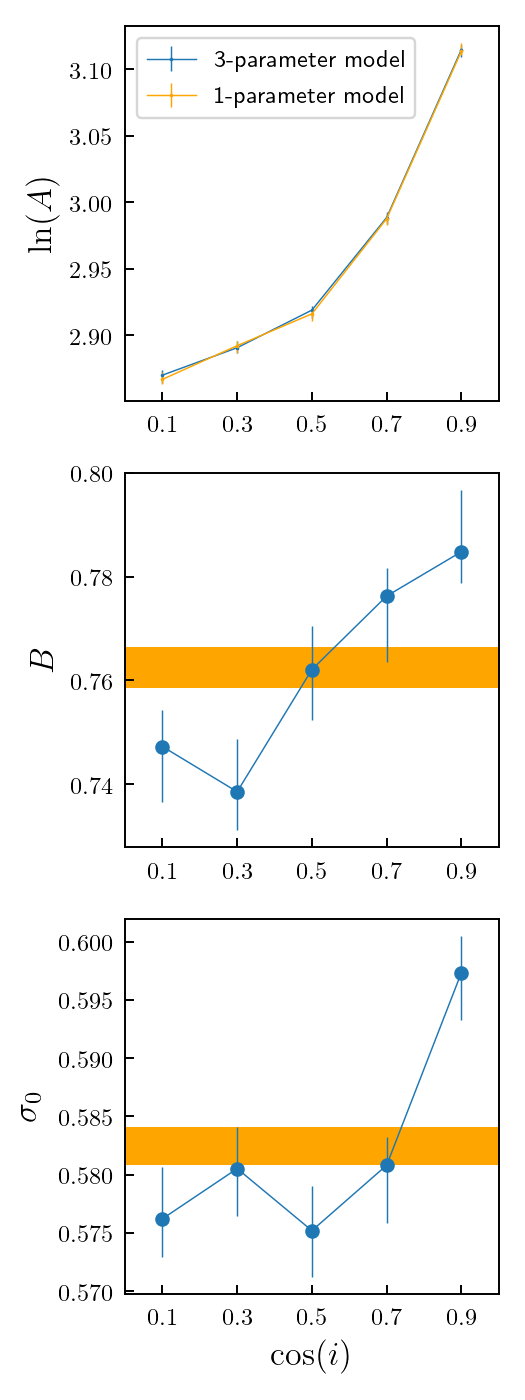}
    \caption{Dependence of redMaPPer richness--mass model parameters on halo orientation $\cos(i)$. Horizontal bands show the mean and 68\% CL range for the global (full-sample) fit for each parameter. 
    The top panel shows best-fit amplitude $\ln(A)$ vs. orientation when the other 2 parameters are allowed to vary with orientation (3-parameter model) and when they are fixed (1-parameter model), indicating little difference. The Bayesian Information Criterion (BIC) test favors the 1-parameter model.
    }
    \label{fig:richness_mass_modelparams}
\end{figure}

Since we have shown that the orientation distribution of redMaPPer-selected clusters is biased, it is important to understand how this may impact the observed cluster richness-mass relation, a key ingredient in cluster cosmology. In this Section, we explore how the cluster richness-mass relation varies with cluster orientation.

Figure \ref{fig:mass_richness_cosi} shows the empirical relation between Buzzard halo mass (defined by $M_{\rm 200m}$) and observed richness for the redMaPPer-matched clusters. 
Following previous work \citep{Saro15, Simet16, Melchior17, McClintock18}, we model the relation between cluster mean richness $\mu(\lambda)$ and halo mass $M$ as a linear relation between $\ln(\lambda)$ and $\ln(M)$, with a pivot point at $10^{14}$~M$_\odot$:
\begin{equation}
\mu(\mathrm{ln} \lambda) = \ln(A) + B\times\big(\mathrm{ln} M - 14\ln(10)\big) ~.
\label{eqn:richness_mass_mean}
\end{equation}
We do not consider the redshift evolution of the richness-mass relation as results from previous multiwavelength scaling relations of galaxy clusters have prescribed a global redshift fit to the richness-mass relation \citep{Simet16} or those that do model the redshift dependence find it consistent with a null dependence \citep{Saro15, Melchior17, McClintock18, Bleem20}. In a recent work, \cite{To21} used Buzzard simulations to quantify the large scale bias of redMaPPer-redMaGic cross correlation that has a redshift dependence at $1-\sigma$ from null and that could be explained by the increase in observed richness at higher redshift from stronger projection effects.

We model the scatter of richness at fixed mass as truncated log-normal scatter that cuts off clusters with $\lambda < 20$:
\begin{equation}
P(\mathrm{ln} \lambda| \mathrm{ln} M) \propto \mathcal{N}(\mu(\mathrm{ln} \lambda), \sigma(\mathrm{ln} \lambda))H(\lambda-20) ~,
\end{equation}
where $H(x)$ is the Heaviside step function.
The variance $\sigma^2$ is the sum of the intrinsic variance $\sigma_0^2$ and a Poisson term due to finite richness, 
\begin{equation}
\sigma^2(\mathrm{ln} \lambda) = \sigma_0^2+ \frac{\mathrm{exp} (\mu(\rm ln\lambda))-1}{\mathrm{exp} ( 2\mu(\rm ln\lambda))}.
\label{eqn:richness_mass_scatter}
\end{equation}

According to Bayes' theorem, the posterior likelihood of the model parameters is given by
\begin{equation}
P(A, B, \sigma_0|\lambda, M) \propto P(\lambda, M|A, B, \sigma_0)  P(A, B,\sigma_0),
\end{equation}
where $P(A,B,\sigma_0)$ is the joint prior on the parameters which we set as non-informative uniform distributions.


The maximum likelihood estimates for the model parameters are found with a Markov Chain Monte-Carlo (MCMC) method implemented through the \textit{pymc} module, assuming uniform priors for $A$, $B$, and $\sigma_0$. We run chains of $10^{6}$ steps for each run, thin them by selecting every $200$ steps, and remove the first $3000$ steps (after thinning) as burn-in, yielding $2000$ steps to sample the posterior distribution. 

The solid line labelled "Combined" in Fig. \ref{fig:mass_richness_cosi} shows the best-fit model to the richness-mass relation for the full redMaPPer sample, with parameters given in the bottom line of Table \ref{tab:richness_mass_bestfit}. The posterior distributions for the "Combined" model parameters shown in Figure \ref{fig:cornerplot_richness_mass} show good convergence of the parameters and minimal correlation among them. The same trends are produced (but not shown) in the posterior distributions for different orientation bins. The reduced chi-square statistics shown in Table \ref{tab:richness_mass_bestfit} show that the model is a good fit to the data. 

Next, we assume that the richness-mass model of Eqns. (\ref{eqn:richness_mass_mean}-\ref{eqn:richness_mass_scatter}) applies separately in each orientation bin. 
The 3-parameter model in each orientation bin is fit independently, with the results shown in Figure \ref{fig:richness_mass_modelparams} and parameter values in the middle box of Table \ref{tab:richness_mass_bestfit}. We find that most of the dependence on orientation comes from the boosting of the amplitude parameter, $\ln(A)$, with $\cos(i)$. We therefore also consider a model in which only $\ln(A)$ varies with orientation, with the other 2 parameters fixed to their global values. The top panel of Fig.  \ref{fig:richness_mass_modelparams} and Table \ref{tab:richness_mass_bestfit} show that this 1-parameter model makes no appreciable change in the best-fit values of $\ln(A)$ in each bin. Moreover, reducing the number of parameters does not significantly compromise the goodness-of-fit of the MLE model relative to the number of extra parameters: as shown in Table \ref{tab:richness_mass_bestfit}, the reduced Bayesian Information Criterion (BIC) for the 1-parameter vs. the 3-parameter model marginally favors the simpler model. 

The best-fit 1-parameter models in each orientation bin are indicated by the dashed lines in Fig. \ref{fig:mass_richness_cosi}: the effect of 
orientation bias on the richness--mass relation is a boost in the amplitude, that is, in observed richness, at fixed halo mass, for halos with major axes aligned with the line of sight. 

While the orientation-bias model studied here captures the behavior of redMaPPer-selected halos in the Buzzard simulations, a caveat is in order before applying the model to redMaPPer-selected clusters in the real universe. In particular, the redMaPPer richness at fixed halo mass in Buzzard has been found to be systematically lower at a $3\sigma$ level from that for redMaPPer clusters with weak-lensing calibrated masses in DES Y1 data \citep{DeRose19} which can be traced to the underestimation of the halo occupation distribution (HOD) of red galaxies identified by the red sequence in Buzzard. If this systematic is relatively independent of richness, we expect our model for the difference in richness amplitude with orientation,  $\Delta \ln(A)$, to retain its validity, even if the central values of $\ln(A)$, $B$ and $\sigma_0$ differ (note that the intrinsic scatter $\sigma_0$ is not constrained in the \cite{McClintock18} weak lensing analysis of DES Y1 clusters). The dependence of the  richness--mass relation on the HOD of red-sequence galaxies can be tested with studies using other simulations, such as the latest cosmoDC2 \citep{Korytov19}, which populates halos with galaxies using a different set of semi-analytic and empirical methods from ADDGALS (Z. Zhang et al., in prep). Alternatively, one can construct and analyze new redMaPPer catalogs from the Buzzard simulations after injecting red-sequence galaxies to match the HOD of DES Y1 data (W. Black et al., in prep).

\begin{table*}
   \begin{center}
   \begin{tabular}{|>{\centering\arraybackslash}m{.08\linewidth}| >{\centering\arraybackslash}m{.08\linewidth}>{\centering\arraybackslash}m{.08\linewidth}>{\centering\arraybackslash}m{.08\linewidth}>{\centering\arraybackslash}m{.08\linewidth}|>{\centering\arraybackslash}m{.08\linewidth}>{\centering\arraybackslash}m{.08\linewidth}>{\centering\arraybackslash}m{.08\linewidth} >{\centering\arraybackslash}m{.08\linewidth}>{\centering\arraybackslash}m{.08\linewidth} | }
        \hline
        \multicolumn{10}{|c|}{Model parameters and BIC for richness--mass template} \\
        \hline
        & \multicolumn{4}{c|}{3-parameter model} & \multicolumn{5}{c|}{1-parameter model} \\
        \hline
        $\cos{i}$& $\ln{A}$  & B & $\sigma_0$ & BIC & $\ln{A}$ & B & $\sigma_0$ & BIC & $\chi^2/\nu$\\
        \hline
        [0.0,0.2) & $2.869 \pm~^{0.004}_{0.006}$ & $0.747 \pm~^{0.007}_{0.011}$ & $0.576 \pm~^{0.004}_{0.003}$ & 8819 & $2.866 \pm~^{0.005}_{0.003}$ &  &  & 8799 & 1.33 \\[3ex]
        [0.2,0.4) &  $2.890 \pm~^{0.005}_{0.004}$ & $0.739 \pm~^{0.010}_{0.007}$ & $0.581 \pm~^{0.004}_{0.004}$ & 8088  & $2.892 \pm~^{0.003}_{0.006}$ & &  & 8064 & 1.26 \\[3ex]
        [0.4,0.6) & $2.919 \pm~^{0.003}_{0.006}$ & $0.762 \pm~^{0.008}_{0.010}$ & $0.575 \pm~^{0.004}_{0.004}$ & 8123
        & $2.916 \pm~^{0.004}_{0.005}$ & \multirow{4}{*}{$0.762 \pm~^{0.005}_{0.003}$} & \multirow{4}{*}{$0.582 \pm~^{0.002}_{0.002}$} & 8104 & 1.20 \\[3ex]
        [0.6,0.8) & $2.988 \pm~^{0.004}_{0.005}$ & $0.776 \pm~^{0.005}_{0.013}$ & $0.581 \pm~^{0.002}_{0.005}$ & 6480 
        & $2.986 \pm~^{0.004}_{0.005}$ &  &  & 6463 & 1.02 \\[3ex]
        [0.8,1.0) & $3.115 \pm~^{0.003}_{0.005}$ & $0.785 \pm~^{0.012}_{0.006}$ & $0.597 \pm~^{0.003}_{0.004}$ & 2648 
        & $3.114 \pm~^{0.006}_{0.003}$ &  &  & 2588 & 0.77 \\[3ex]
        All & \multicolumn{4}{c|}{NA} & $2.956 \pm~^{0.003}_{0.001}$  &  &  & 29807 & 1.09 \\[3ex]
        \hline
    \end{tabular}
    \end{center}
    \caption{Maximum Likelihood estimates and 68\% CL errors of richness-mass model parameters for redMaPPer clusters as a function of halo orientation $\cos(i)$ and for the full cluster sample ("All"). The middle box shows results when all 3 model parameters are allowed to vary with $\cos(i)$ (3-parameter model); right-most box shows results when only $\ln(A)$ is allowed to vary (1-parameter model). Also shown are the Bayesian Information Criterion (BIC) values for each case; the slightly lower values for the 1-parameter model indicate that it is marginally preferred. The reduced chi-square statistics $\chi^2/\nu \sim 1$ show that the 1-parameter model is a good fit to the data. 
    }
    \label{tab:richness_mass_bestfit}
\end{table*}

\section{Correlation of triaxiality with other systematics}

Orientation bias is one significant systematic for the cluster richness-mass relation; miscentering and projection effects are two others. In modeling these systematics for cluster cosmology, it is important to know the degree to which they may be correlated. In this Section, we explore possible correlation of orientation bias with the other two.

\subsection{Miscentering}
\label{subsec:miscentering}
As noted above in Section \ref{sec:cl_sel}, in the simulated cluster catalog 37\% of the matched clusters are miscentered in the sense that the galaxy identified by redMapper as the BCG is not the central galaxy in the corresponding Buzzard halo. In both the simulation and the real universe, miscentering can happen for a number of reasons. 
For example, a recent halo merger may result in two nearly-central galaxies of comparable luminosity, or a recent burst of star formation may move the central galaxy's color off the locus of the red sequence. 
\citep{Cooke19, Ragone-Figueroa20, Zenteno20}. 
Alternatively, a red foreground galaxy along the line of sight to a cluster may be misidentified as the BCG, although 
Section \ref{subsec:projection} indicates that this is rare in the Buzzard simulations.


The miscentering distribution for redMapper clusters in DES Y1 data was estimated through comparison of redMaPPer BCG angular positions with the peaks of X-ray emission for a subsample of clusters with Chandra archival data \citep{Y.Zhang19}. A number of studies have indicated that X-ray peaks are accurate proxies for the centers of cluster potential wells, though they are subject to systematic errors as well
\citep{LinMohr04, Song12, Stott12, Mahdavi13, Lauer14}. 
In \cite{Y.Zhang19}, based on 144 redMaPPer clusters with X-ray data, $75\pm8$\% of the redMapper clusters were found to be centered, i.e., they have very small projected separation between redMaPPer BCG and X-ray centroid. For the remainder, 
the distribution of radial separation between redMaPPer BCGs and X-ray peaks was modeled as a sum of 
a declining exponential and a gamma function. 


Here, we study the distribution of projected separation, $R_{\rm sep}$, between redMaPPer BCGs and Buzzard central galaxies for halo-matched clusters in the simulation. Since the separation is expected to scale with cluster size, we use the scaled separation, $R_{\rm sep}/R_\lambda$, where $R_\lambda = 1 \text{$h^{-1}$~Mpc}(\lambda/100)^{0.2}$ is the characteristic circular aperture for the redMaPPer cluster finder. 

We note here the difference in definition between centers. In real data the centering property for a single cluster is not known. Rather the separation distance between optical and X-ray center is modeled as a joint distribution for centered and miscentered clusters with the centered fraction as a model parameter with a maximum likelihood of $75\% \pm 8\%$. In the simulations the centering of each individual cluster is a known quantity determined by whether the central galaxy determined by redMaPPer and the halo are one and the same. Among the 23658 halo-matched clusters, 14905 were correctly centered and 8753 are miscentered, the centered fraction being $63\%$ which is within  $2\sigma$ the centering fraction using X-ray follow-up \citep{Y.Zhang19}.

In contrast, Buzzard populates halo centers with galaxies using the ADDGALS algorithm, and by construction a galaxy lies in the halo center. For redMaPPer clusters correctly centered on the halo central galaxy their coordinates are perfectly matched. We define the distance between the redMaPPer chosen BCG and the true halo center as the miscentering separation distance $R_{\rm sep}$.

\begin{figure}
	\centering
	\includegraphics[width=0.5\textwidth]{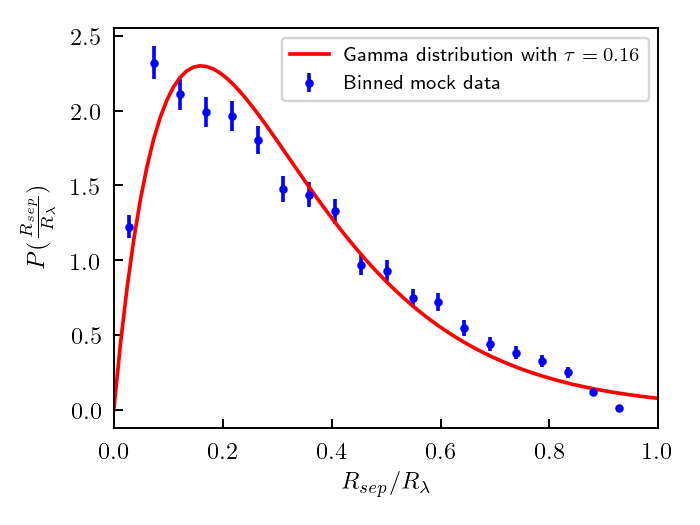}
    \caption{Probability distribution of the projected separation between Buzzard halo central galaxies and redMaPPer BCGs in the miscentered population. Scatter plots are the binned mock data points with Poisson error and the line is the best fit Gamma distribution. The two distributions are consistent according to the Kolmogorov-Smirnov test at a $\alpha=0.05$ significance level.}
    \label{fig:miscentered_dist}
\end{figure}

The resulting separation distribution is shown in Fig. \ref{fig:miscentered_dist}; the distribution is peaked at $R_{\rm sep} = 0.1 R_\lambda$, with a tail that extends to  $R_{\rm sep} \simeq R_\lambda$. The shape of the distribution is well-fit by a $\Gamma$ distribution of functional form
\begin{equation}
    P_{\rm miscent}(x|\tau) = \frac{x}{\tau^2}\exp{\big(-x/\tau\big)},
\end{equation}
where $x \equiv R_{\rm sep}/R_{\rm \lambda}$. Using methods of least squares, the best fit characteristic scale is found to be $\tau = 0.16$ which is well within the $1-\sigma$ range of the characteristic scale for Chandra to DES center offset found in \cite{Y.Zhang19}. Using the Kolmogorov-Smirnov test, we find that the binned dataset is consistent with the best-fit Gamma distribution at a $\alpha=0.05$ significance level. 

\begin{figure}
	\centering
	\includegraphics[width=0.45\textwidth]{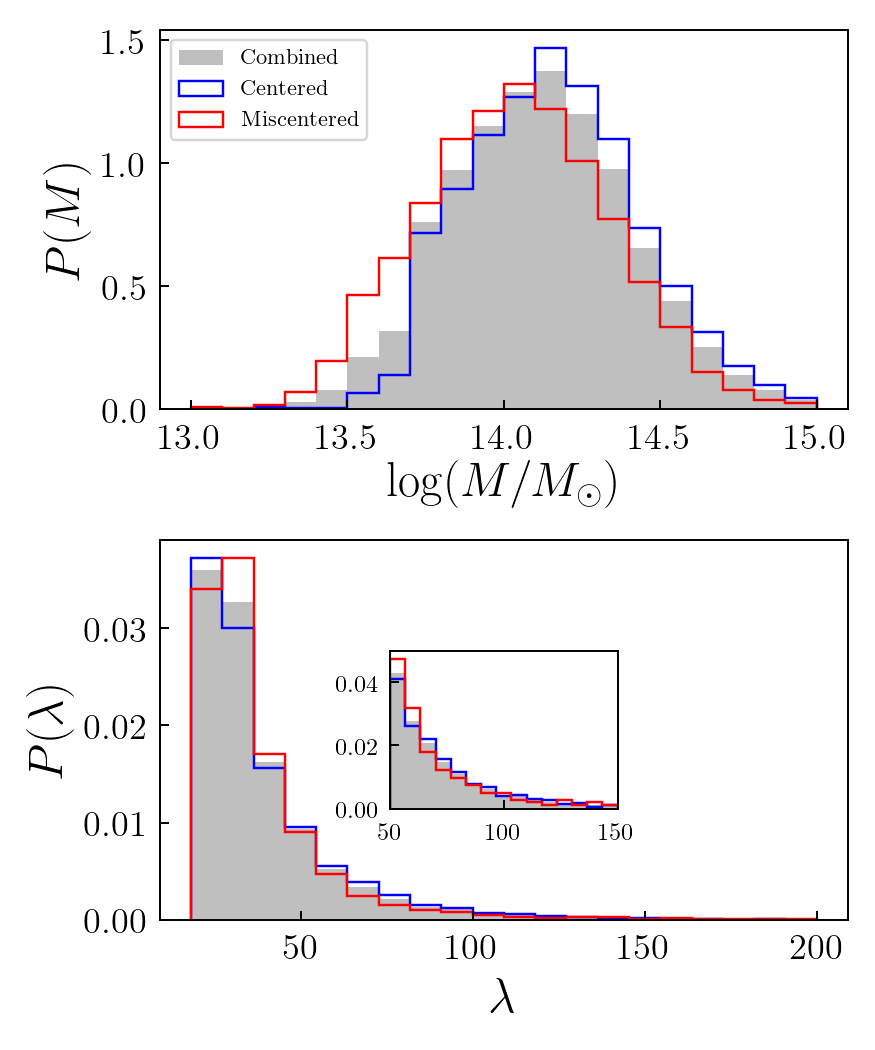}
    \caption{{\it Upper panel:} Mass distribution of the centered and miscentered redMaPPer clusters in the Buzzard simulations. The centered population is peaked at a higher mass.  {\it Lower panel:} Richness distributions of centered and miscentered clusters and for the entire cluster sample. The inset plot shows a slightly higher fraction of centered clusters at high richness. 
    }
    \label{fig:miscentered_mass_richness_distribution}
\end{figure}

We study differences in the properties of the centered and miscentered cluster populations in the simulation in Fig. \ref{fig:miscentered_mass_richness_distribution}. The upper panel shows that the probability distribution of cluster mass for the centered population is peaked at a slightly higher mass than for the miscentered population, that is, it is the lower-mass clusters that tend to be miscentered, which suggests that this may be a mass dependent bias more prone to low-mass and low-richness clusters. The same trend was not observed with X-ray luminosity and temperature, variables sensitive to the cluster mass with a sample size of only 144 redMaPPer SDSS clusters with X-ray follow up \citep{Y.Zhang19}. Near-future X-ray surveys as eRosita \citep{Hofmann17}, which aims to detect $10^5$ clusters with a lower mass limit of $\sim 10^{14}~M_{\odot}$, will provide a much better handle on the mass distribution of centered and miscentered clusters. The lower panel of Fig. \ref{fig:miscentered_mass_richness_distribution} shows that the normalized richness distribution of the centered clusters is higher than that of the miscentered ones at $\lambda > 60$, though the difference is marginal.


The centered fraction increases with increasing richness, from 63\% for the full sample ($\lambda>20$) to 60\% for $\lambda > 40$, 67\% for $\lambda > 60$ and 69\% for $\lambda > 80$. This trend is qualitatively consistent with the consistency test carried out on data by \cite{Y.Zhang19}: they compared redMaPPer BCG positions for DES and SDSS clusters where the two data sets overlap and found that for $\lambda > 40$ a large fraction of the BCG positions were within $0.05 R_\lambda$ of each other. The archival data from XMM and Chandra has a sharp richness cutoff of $\lambda \gtrapprox 70$ \citep{Farahi19}, so any trend of miscentering of BCGs relative to X-ray centroids with richness is not yet detectable with current data. 

To quantify the impact of miscentering on the redMaPPer richness estimate in the Buzzard simulations, we consider two approaches. The first method is to recalculate the observed richness by assigning the cluster center onto a different galaxy. It has the advantage that it can also be applied to cluster data but the disadvantage that it involves additional assumptions that have not been fully tested. For each cluster, the redMaPPer algorithm initially identifies five galaxies as candidates for the BCG. At the end of its iterative procedure, it assigns a final probability of being the BCG to each of these five, produces richness estimates, $\lambda_i$, $i=1,...,5$ assuming each of them is the BCG, and identifies the most probable as the BCG, with corresponding richness estimate $\lambda_i$. As the probability of it being the true center drops for each candidate, a comparison of richness for clusters targeted at different central candidates would yield information on the potential degree of miscentering for each cluster. 

\begin{figure*}
    \subfigure{\includegraphics[width=0.45\textwidth]{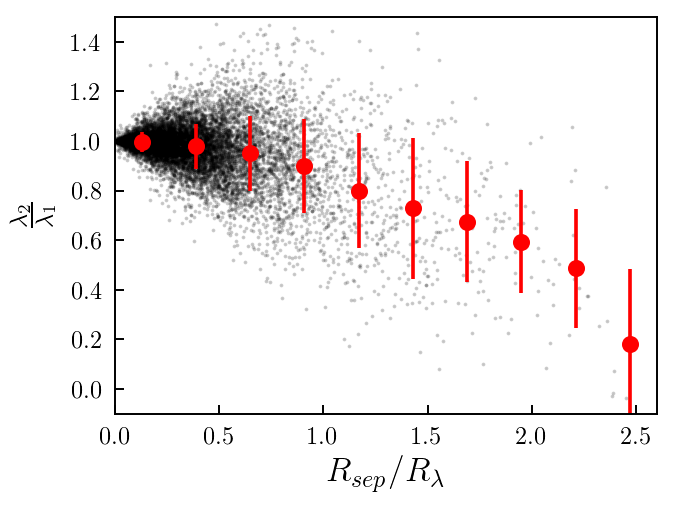}}
    \subfigure{\includegraphics[width=0.45\textwidth]{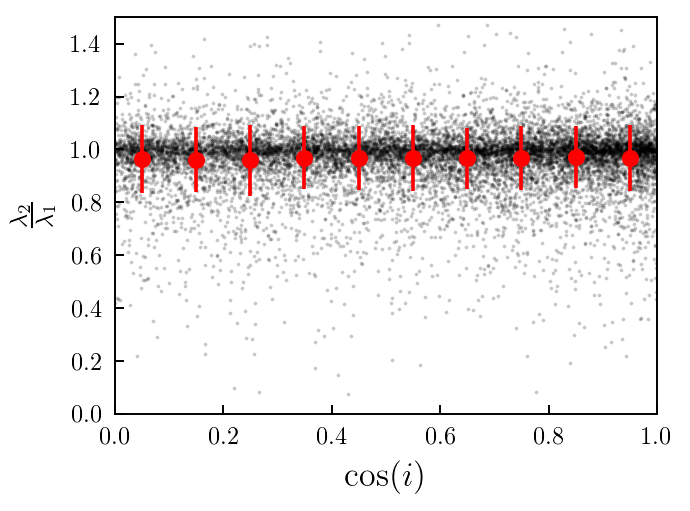}}
    \\
    \subfigure{\includegraphics[width=0.45\textwidth]{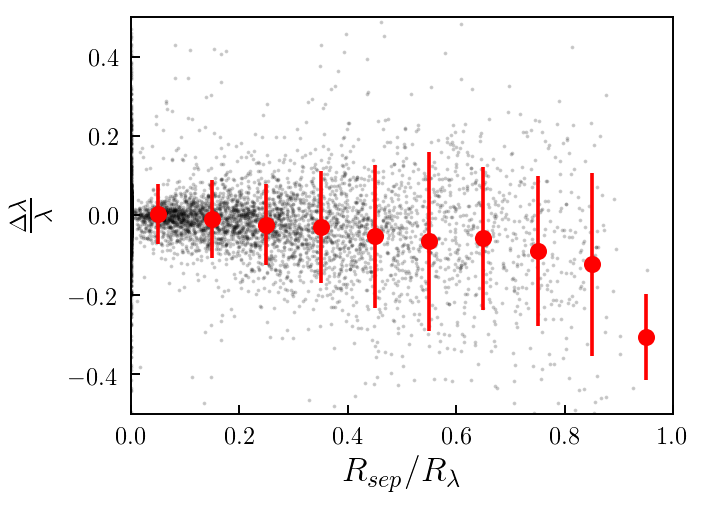}}
    \subfigure{\includegraphics[width=0.45\textwidth]{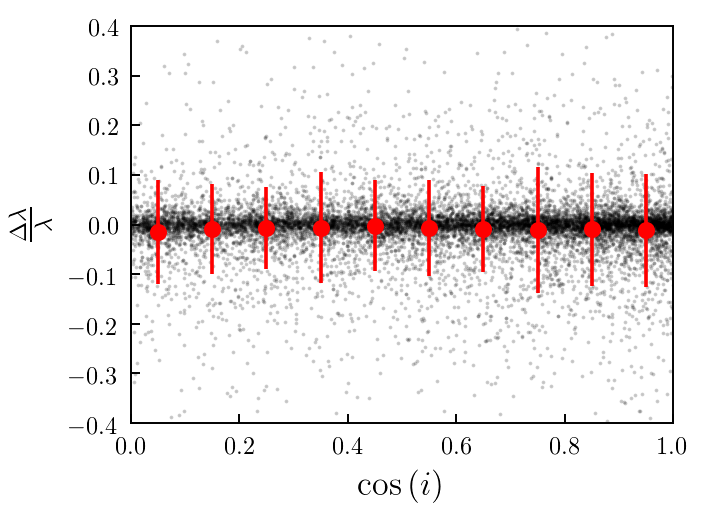}}
    
    \caption{{\it Left panels:} Richness bias vs. miscentering separation for redMaPPer clusters in the Buzzard simulation. Both richness bias metrics $\lambda_2/\lambda_1$ and  $\Delta \lambda/\lambda$ show larger bias and increased dispersion at larger miscentering distance. 
    {\it Right panels:} Richness bias vs. orientation. The mean values of the richness bias metrics show no correlation with halo orientation angle, $\cos(i)$.
    }
    \label{fig:miscentering_vs_cosi}
\end{figure*}

In this first approach, we can quantify the bias in miscentering by taking the ratio of the richness centered at the second most probable galaxy to the first most probable cluster central galaxy among the 5 candidates identified by redMaPPer. This ratio $\lambda_2/\lambda_1$ is an indication of the potential bias in observed richness that miscentering could play when choosing a different cluster center. Among the many selection effects of redMaPPer that come into play in the measurement of this quantity, it is primary a function of the separation distance between the two central candidates---$\lambda_2/\lambda_1$ shifts downward from unity with increasing separation distance $R_{\rm RM\_ sep}$ between the cluster candidates, and also notably so does the dispersion increase with $R_{\rm RM\_ sep}$. Here $R_{\rm RM\_ sep}$ is the separation distance between the two redMaPPer central candidates which in some clusters could be the halo-cluster separation distance $R_{\rm sep}$ but is often not the case. As shown in the left panels of Fig. \ref{fig:miscentering_vs_cosi} and in Fig. \ref{fig:miscentered_dist}, $R_{\rm sep}$ goes out to $\sim 1 R_{\lambda}$ while $R_{\rm RM\_ sep}$ can be extended to $\sim 2.5 R_{\lambda}$.  

The second method of quantifying the impact of miscentering on richness gives a ``ground-truth" estimate of the richness bias, but it can only be estimated in the simulation, not from observations. There is a version of the redMapper catalog for the Buzzard simulation, called the \textit{halorun} catalog, in which the redMaPPer BCG is constrained to be the halo central galaxy for each halo-matched cluster. By construction, correctly centered clusters in the \textit{fullrun} redMaPPer catalog that we have been discussing so far have the same richness as those in the \textit{halorun} catalog. On the other hand, for the miscentered \textit{fullrun} clusters, there is a bias in the estimated richness due to miscentering characterized by 
\beqa
\frac{\Delta\lambda}{\lambda} = \frac{\lambda_{\rm fullrun}-\lambda_{\rm halorun}}{\lambda_{\rm fullrun}} ~.
\eeqa
This fractional shift in richness is plotted as a function of the scaled miscentering separation in the lower left panel of Fig. \ref{fig:miscentering_vs_cosi}.

It is apparent from visual inspection in the left panels of Fig. \ref{fig:miscentering_vs_cosi} that both methods of quantifying miscentering bias that richness bias increases in amplitude and dispersion with scaled separation as has been shown using DES Y1 clusters with X-ray follow-up data. 

Having shown that the miscentering properties of the Buzzard redMaPPer catalog are consistent with those in DES Y1 data, we now turn to examining whether miscentering and triaxiality are correlated systematics. We do this by measuring the miscentering bias as a function of halo orientation, using both of the metrics described above. As the right panels of Fig. \ref{fig:miscentering_vs_cosi} show, we find that the mean values and dispersion of the two metrics have no systematic dependence on $\cos(i)$. Miscentering and triaxiality can thus be treated as independent systematics. 

The fact that we find no correlation between these two systematics is useful for the modeling of systematics in future weak lensing studies but should not come as too unexpected in light of their different physical origins.  Miscentering occurs when mergers introduce identical central galaxy candidates or from the star formation properties of the central galaxy that shifts its color out of the red-sequence \citep{Cooke19, Ragone-Figueroa20, Zenteno20}, effects completely different from the geometric boosting in richness when clusters are oriented along the line of sight that induce triaxiality bias. 

We also test if miscentering can be attributed to line of sight projections whose effect on clusters we describe in detail in subsection \ref{subsec:projection}. If miscentering is due to projection effects then the BCG at the center of the matched-halo would be of a different redshift and not belong as a member of the matched redMaPPer cluster. Within the allowed $\Delta z\pm0.05$ redshift separation between halo and cluster in our matching algorithm, all of the BCGs at the halo center belong as a member of the matched redMaPPer cluster. Additional tests beyond the scope of this paper need to be conducted to in order to conclude whether miscentering can be attributed to projection effects and if so to what degree, but simulations from Buzzard suggests that this may not be a strong effect.

\subsection{Projection}
\label{subsec:projection}
In this section we test for correlations between triaxiality and projection effects. Projections effects were modeled and quantified in \cite{Costanzi19} using a different Buzzard halo catalog populated according to the assigned "true" richness--mass relation of \cite{Simet16}, and adopts an empirically calibrated back/foreground contamination to account for projection effects on the observed richness. We denote this catalog as the \textit{C19 projection catalog}. Below we summarize the properties of projection effects and the quantities used in the C19 projection mock catalog for our analysis.

Cluster richness suffers from projection effects when non-member galaxies along the line of sight to a cluster are mistakenly classified as cluster members. These may be randomly located galaxies along the line of sight, galaxies spatially correlated with the cluster due to large-scale (e.g., filamentary) structure, or galaxies in a lower-richness cluster along the line of sight that ``leak'' into a larger one, a process in redMaPPer known as \textit{percolation} \citep{Costanzi19}. In combination, they bias the observed richness $\lambda^{\rm obs}$ away from the true richness $\lambda^{\rm true}$ by the amount:
\beqa
\lambda^{\rm obs} - \lambda^{\rm true} = \Delta^{\rm bkg} + \Delta^{\rm prj}_{\rm non-cor} + \Delta^{\rm prj}_{\rm LSS} + \Delta^{\rm prc}
\label{eqn: proj_prob}
\eeqa 

Each component contributes to the observed richness in a different form. Background scatter, $\Delta^{\rm bkg}$, is assumed to be normally distributed around the true richness. The sum of the projection terms due to non-correlated clusters, $\Delta^{\rm prj}_{\rm non-cor}$, and correlated large-scale structure, $\Delta^{\rm prj}_{\rm LSS}$, are modeled as an exponential function with a cutoff at $\Delta^{\rm prj} \geqslant 0$, to ensure an upscatter of $\lambda_{\rm obs}$ as is physically motivated. The observed richness is painted on in the mock catalog by summing the richness of clusters along the light of sight weighted by the redshift kernel $w(\Delta z, z)$:

\beqa
\lambda_i^{\rm obs} = \lambda_i^{\rm true} + \Delta_i^{\rm prj} = \lambda_i^{\rm true} + \sum^N_{j\neq i} \lambda_j^{\rm true}f^{\rm A}_{ij}w(\Delta z_{ij}, z_j),
\eeqa
where $f^A_{ij}$ is the geometric masking fraction of object $j$ over $i$ for an object $j$ that's (partially) in the line of sight of $i$, and $w(\Delta z_{ij}, z_j)$ the redshift kernel which, as a function of redshift of $i$ and the redshift separation between $i$ and $j$ is modeled as the functional form:
\begin{equation}
        w(\Delta z | z_{\rm cl}) =
        \Bigg\{ \begin{array}{rl}
            1 - \frac{(\Delta z)^2}{\sigma_z(z_{\rm cl})^2} , & \lvert\Delta z\rvert < \sigma_z(z_{\rm cl}) \\
            0, & \text{otherwise},
        \end{array} 
    \end{equation}
which can be interpreted intuitively as the diminishing strength of projection effects with redshift separation $\lvert \Delta z \rvert$ up to a maximum separation of $\sigma_z(z_{\rm cl})$.

For each cluster, its $\sigma_z^{\rm cl} (z)$ is fitted by sliding the redMaPPer redshift center away from the true cluster redshift so as to remove the excess richness $\Delta^{\rm prj}$ due to projection as a function of the redshift separation between assigned and true redMaPPer redshift. To recover the ``leakage" function for clean line of sights, \cite{Costanzi19} chooses the lower $5\%$ of clusters in a given redshift as the leakage function. It is fit with a piecewise log-linear model with a transition at $z=0.32$. Data from SDSS redMaPPer clusters \citep{Costanzi19} show that at $z \lesssim 0.3$ projections  are from the width of the red-sequence and increase monotonically with increasing redshift from increasing photometric errors. At $z \gtrsim 0.3$ projection effects flatten out as the SDSS survey is no longer volume limited but magnitude limited, the faintest cluster galaxies residing near the magnitude limit of the survey at redshift above $0.3$.

In this paper we introduce the derived quantity 
\beqa
\log{\big(\sigma_z^{\rm proxy}(z_{\rm cl})\big)} =  \log{\big(\sigma_z^{\rm cl}(z_{\rm cl})\big)} - \log{\big(\sigma_z^{\rm 5\%}(z_{\rm cl})\big)}
\label{eqn:sigma_z}
\eeqa
as the difference between the log-scaled $\sigma_z$ of an individual cluster and the lower $5\%$ envelope of $\sigma_z$ for all clusters at the redshift bin of the cluster. This quantity $\sigma_z^{\rm proxy} (z_{\rm cl})$ can be seen as the level of intrinsic excess projection after eliminating background noise and redshift-dependent observational biases.

Percolation is added into the full model of projection when clusters of lower richness are ``absorbed" into one with higher richness. For each cluster $j$ with richness smaller than that of $i$, the richness is taken from $j$ to $i$ by the amount
\beqa
\Delta_{\rm j}^{\rm prc} = \sum^N_{j < i} \lambda_j^{\rm true}\big(1-f^{\rm A}_{ij}w(\Delta z_{ij}, z_j)\big),
\eeqa
whose probability distribution $P(\Delta^{\rm prc}|\lambda^{\rm true}, z)$ is empirically determined to well resemble a boxcar function with $\Delta^{\rm prc}  \in [-\lambda^{\rm true}, 0]$. 

In the C19 projection catalog, each cluster is assigned a true richness using an empirically calibrated richness--mass relation from \cite{Simet16} and given an observed richness using the projection effect algorithm described above by way of the redshift kernel $w(\Delta z | z_{cl})$. Hence the difference between the true and observed richness in this mock is due to projection effects alone. The probability distribution for $P(\Delta|\lambda^{\rm true}, z)$ for each component is then fit using this C19 projection mock, and upon convolution of the probability distributions for each individual component in Equation \ref{eqn: proj_prob} we arrive at the final expression for $P(\lambda^{\rm obs}|\lambda^{\rm true}, z)$. We refer the reader to \cite{Costanzi19} for the full expression and best-fit parameters. All halos in the mock projection catalog are artificially assigned an observed and true richness, whether or not such a halo could be detected and matched to a redMaPPer cluster. The observed richness is thus biased only from projection effects and does not suffer from all the other selection effects, including triaxiality and miscentering, that would exist had the halos undergone redMaPPer detection and cluster matching. This technique effectively isolates projection effects from potentially correlated systematics in the same vein that we used the \textit{halorun} catalog to isolate miscentering effects. 

\begin{figure}
\centering
\subfigure[Correlation of $\sigma_z^{\rm proxy}$ with halo orientation.]{\includegraphics[width=0.4\textwidth]{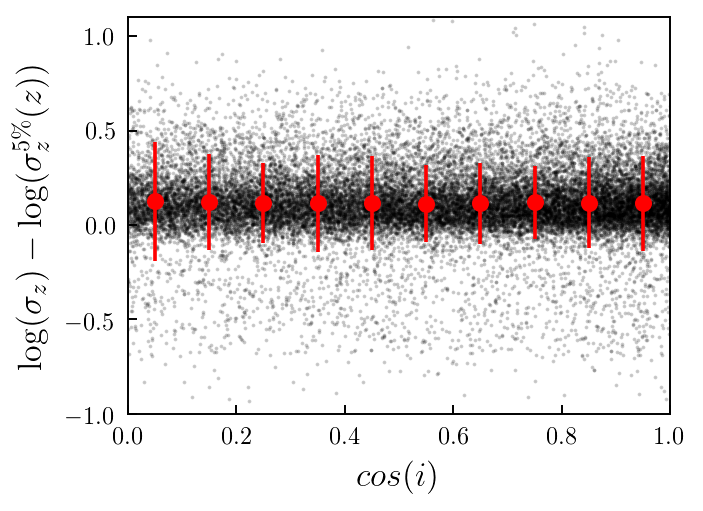}\label{subfig:sigmaz_proj_cosi}}
\subfigure[Correlation of $\frac{\Delta\lambda}{\lambda}$ with halo orientation.]{\includegraphics[width=0.4\textwidth]{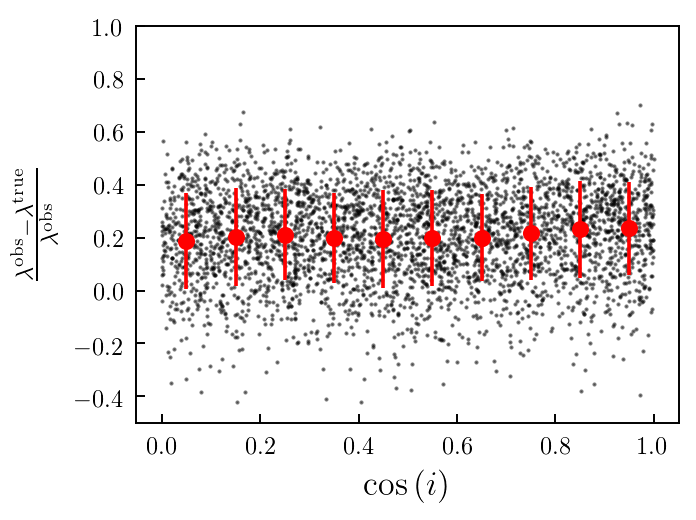}\label{subfig:dlmda_proj_cosi}}
\caption{Correlation of projection strengths and halo orientations measured in two mock catalogs. Top panel shows the measurement in the Buzzard simulations, where the $\sigma_z^{\rm proxy}$ (defined in Equation \ref{eqn:sigma_z}) is used to estimate the strength of projection effects.  Bottom panel shows the measurement in the C19 projection mock, which is constructed using the same halo catalog as the Buzzard simulations. In the C19 catalog the galaxies are populated using a richness--mass relation and the observed richness is generated using a semi-analytic model (described in Section \ref{subsec:projection}).  In this mock, because we know the true galaxy content in each halo, we use the fractional difference between the observed richness and true richness (defined in Equation \ref{eqn:fracdiff_richness_proj}) as a proxy for projection. In both panels, we find there is no correlation between projection strengths and halo orientations. 
}
\label{fig:projection_vs_cosi}
\end{figure}

\begin{figure}
	\centering
	\includegraphics[width=0.45\textwidth]{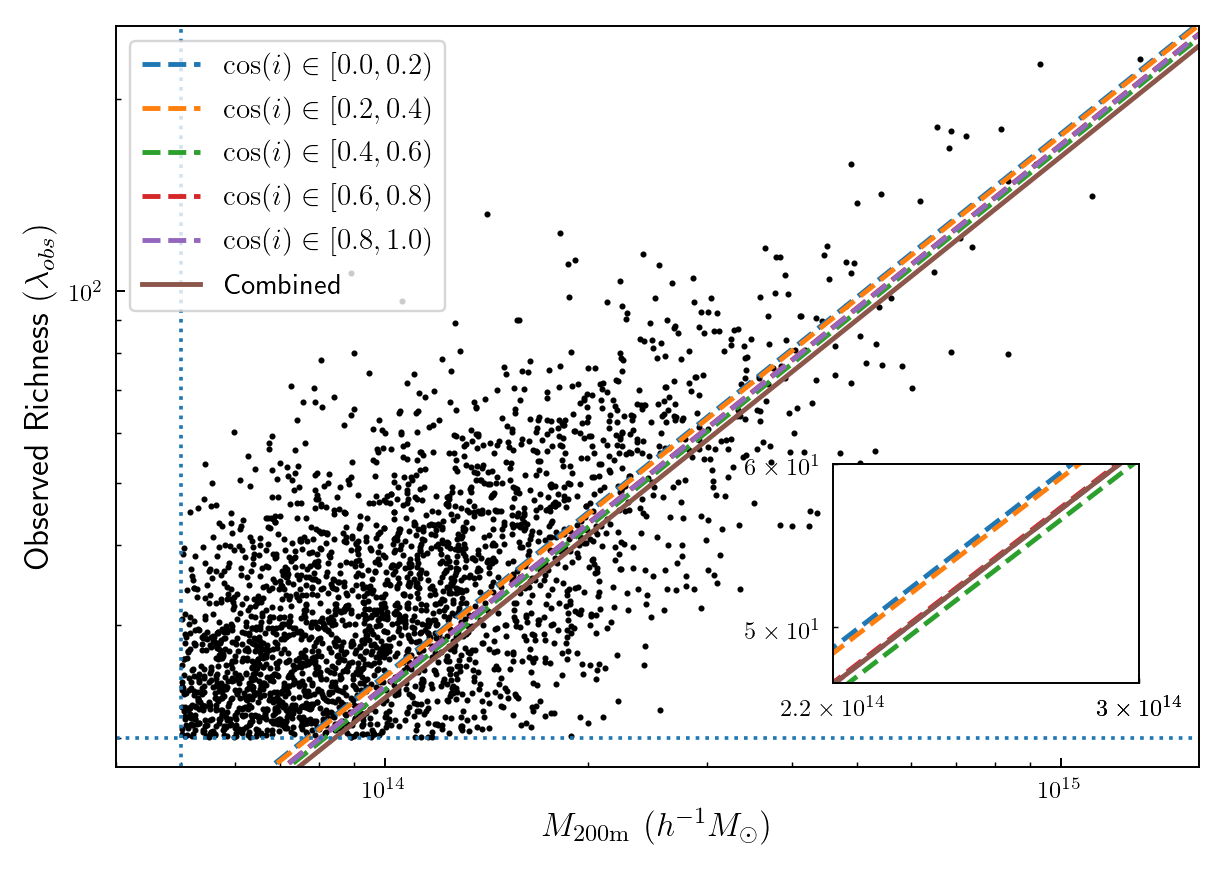}
    \caption{Observed richness--mass relation for different orientation bins in the projection mock catalog. No difference is observed in the observed richness--mass relation in the projection catalog with clusters of different orientation bins. The $1-\sigma$ contours for the best fit parameters $\ln{(A)}$, $B$ and $\sigma_0$ (not shown) in all bins closely overlap with one another, indicating no correlation between the two systematics. The dashed horizontal line indicates the richness cut at $\lambda >20$ and dashed vertical line the mass cut at $M_{\rm 200m} > 5\times 10^{13}~h^{-1}M_{\odot}$.}
    \label{fig:projection_richness_mass}
\end{figure}

We find that projection effects are independent from triaxiality. Figure \ref{subfig:sigmaz_proj_cosi} shows that $\sigma^{\rm proxy}$, the strength of projection effects due to large-scale structure, is not correlated with $\cos(i)$. We further inspect the full scope of projection effects by studying the fractional difference between the observed and true richness in the projection mock catalog,
\begin{equation}
\frac{\Delta \lambda_{\rm prj}}{\lambda_{\rm prj}} = \frac{\lambda_{\rm prj}^{\rm obs} -\lambda^{\rm true}}{\lambda_{\rm prj}^{\rm obs}},
\label{eqn:fracdiff_richness_proj}
\end{equation}
which shows no correlation with $\cos{(i)}$, as shown in Figure \ref{subfig:dlmda_proj_cosi}. Finally, we run our fit to the richness--mass relation in the projection catalog of $\lambda^{\rm obs}_{\rm prj}$ binned in $\cos(i)$ and observe no difference in the observed richness--mass relation, shown in Figure \ref{fig:projection_richness_mass}. The $1-\sigma$ range of the best-fit parameters for the log-linear richness--mass template between different $\cos(i)$ bins all closely overlap with each other, with no clear trend. 

The lack of correlation between projection and orientation may be puzzling at first in light of a common physical origin of these effects. The $\rm{\Lambda CDM}$ model of hierarchical structure formation facilitates the preferential gravitational collapse of dark matter halos that become galaxy clusters along the nodes of large-scale filaments. It is also widely understood that a halo's semi-major axis is preferentially aligned with the direction of the associated filament for halos residing in over-densities (e.g. \cite{Hahn07b}, \cite{Forero-Romero14}). It is thus sensible to expect a correlation between the strength of projection effects and halo orientation for halos residing in filaments. 

The lack of correlation can be explained by the stochasticity of these effects along with the fact that not all halos share the same physical origin for this set of systematics. The boosting in richness from projection is from uncorrelated background noise and correlated large scale structure, the latter playing a much larger role. Adding the large scale structure into the modeling of the observed richness for projection boost the richness perturbation $\Delta^{\rm prj}$ by a factor of 2 and 4 in the $\lambda^{\rm true}$ range $20-100$ \citep{Costanzi19}. It is also observed by N-body simulations from \cite{Sunayama20} that a minority of clusters that reside in large scale filaments is responsible for the boosting of the stacked weak lensing signal of halos (see Section \ref{sec:weak_leasing} on weak lensing) due to projection effects. This set of studies suggests that a small batch of clusters is responsible for the large degree of bias from projection effects. 

Triaxiality bias, on the other hand, can occur whether halos reside in large scale filaments or in voids. That all halos, regardless of its external environment, is subject to the same degree of triaxiality bias while not the case for projection bias would explain the lack of correlation among an ensemble of stacked clusters. It would be interesting as a follow-up study to know if the correlation between projection and triaxiality can be detected for the minority of clusters residing in large scale structures that heavily boost the projection observable, but for the purposes of modeling redMaPPer selection effects, it is sufficient to know that for the entire sample of $\lambda^{\rm obs} > 20$ clusters detectable by redMaPPer, projection and triaxiality can be treated as separate systematics. A further study using spectroscopic redshift measurements of redMaPPer member galaxies from Magellan telescope data (Gruen D., in prep.) will provide the shape and orientation of clusters as well as test for non-member galaxies projected along the line of sight misidentified by redMaPPer, serving as a follow-up test of the correlation of these systematics using real data. 

\section{Effect of halo orientation on weak lensing profile}
\label{sec:weak_leasing}
The effects of triaxiality on cluster optical detection are twofold---one through the boosting of the richness-mass relation as was covered in Section \ref{sec:richness--mass}, the other through the boosting of radially dependent weak lensing signals.

This section quantifies the latter effect. It is split into three subsections---Section \ref{subsec:delta_sigma_profile} models the boosting effect of the cluster weak lensing signal in the Buzzard simulations for individual halos before applying the redMaPPer cluster finder; Section \ref{subsec:delta_sigma_bias_redM_selected} combines the result from Section \ref{subsec:delta_sigma_profile} and our richness-mass model from Section \ref{sec:richness--mass} to predict the observed boosting in stacked cluster lensing profiles at different richness bins after redMaPPer selection; Section \ref{subsec:mass_bias} uses the result from Section \ref{subsec:delta_sigma_bias_redM_selected} to conduct a Fisher matrix forecast on the mass bias of triaxiality for redMaPPer clusters stacked in different richness bins.

\subsection{Modeling the effects of halo orientation on excess surface density before redMaPPer selection}
\label{subsec:delta_sigma_profile}
In this section, we measure the excess surface densities of all halos with convergent shape measurements in a lightcone of $z<0.90$. The masses of halos are binned in mass bins of $[5\times10^{13},~10^{14}),~ [10^{14},~2\times10^{14}),~[2\times10^{14},~4\times10^{14})~\rm{and}~ [4\times10^{14},~\infty)$~$h^{-1}M_{\odot}$, and redshift bins of $[0,~0.34),~[0.34,~0.5),~[0.5,~0.7) \rm~and~[0.7,~0.9)$, for a total of 16 bins.

Another common expression for the density inside a halo is the halo--matter correlation $\xi_{hm}(r)$, which is related to the surface density $\Sigma$ through the relation
\beqa
\Sigma(R) = \overline{\rho}_m\int^{+\infty}_{-\infty}\Bigg(1+\xi_{hm}\Big(r = \sqrt{R^2 + z^2}\Big)\Bigg) dz,
\label{eqn:halo_matter_corr}
\eeqa
where $\overline{\rho}_m$ is the mean matter density at the redshift of the cluster, $R$ is the projected radius in the plane of the sky, and $z$ is the length along the line of sight. 

In weak lensing, the tangential shear $\gamma_t$ of the galaxies relative to the center of each foreground halo is related to the excess surface density by the relation
\beqa
\Sigma_{\rm crit} \gamma_t = \overline{\Sigma}(<R) - \Sigma(R) \equiv \Delta\Sigma(R),
\label{eqn:DeltaSigma}
\eeqa
where the critical surface density $\Sigma_{\rm crit}$ defined as
\beqa
\Delta\Sigma_{\rm crit} = \frac{c^2}{4\pi G}\frac{D_s}{D_l D_{ls}},
\eeqa
and where $D_s$, $D_l$ and $D_{ls}$ refer to the angular diameter distances to the source, to the lens, and between the lens and source, respectively. 

In this paper we measure $\DS(R)$, which has a one-to-one relationship with $\Sigma(R)$ and $\gamma_t$, all of which can be determined from the underlying halo--matter correlation $\xi_{hm}(r)$ and a fiducial cosmology for determining $\Sigma_{\rm crit}$. In the following sections, in order to reduce the clutter in the equations for modelling excess surface density as a function of orientation we use $\mu$ as a shorthand for $\cos(i)$. 

When we measure $\Delta\Sigma(R)$ from the simulations, we use  projected radii $R$ extending from $0.1~h^{-1}$~Mpc to $100~h^{-1}$~Mpc in 30 equally log-spaced bins, and a projected distance symmetric about the halo of $\Delta D_p = 10,~50,~100,~200~h^{-1}$~Mpc. For ease of visualization, the orientation dependence is plotted and fitted onto a template as the quantity 
\beqa
F(R, \mu) = \log\frac{\Delta\Sigma(R, \mu)}{\overline{\Delta\Sigma}(R)},
\label{eqn:DS_ratio}
\eeqa
where $\Delta\Sigma(R, \mu)$ is the average profile in an orientation bin for a given mass and redshift bin, and $\overline{\Delta\Sigma}(R)$ is the averaged profile across all orientation bins in the same mass and redshift bin.

\begin{figure}
	\centering
	\includegraphics[width=0.45\textwidth]{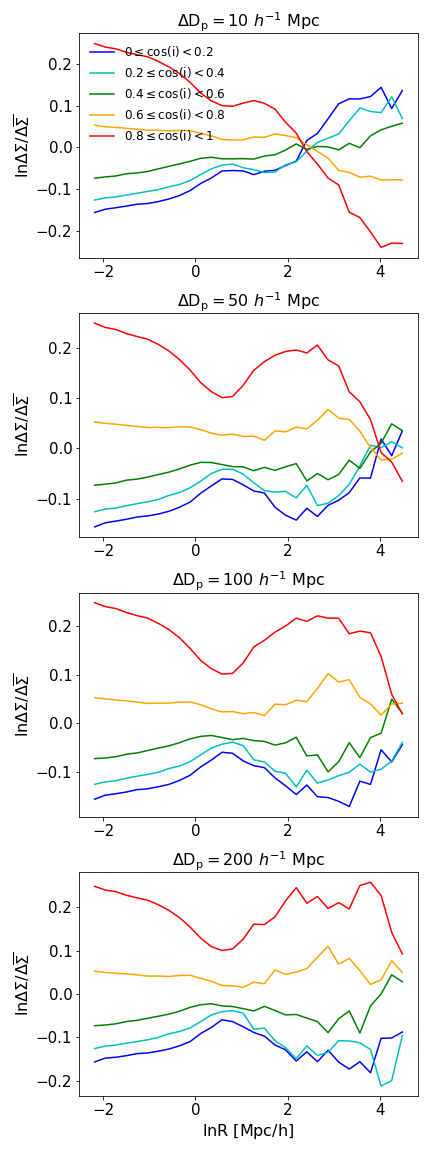}
    \caption{$\Delta\Sigma(R,\mu)$ for $M \in [10^{14}, 5\times10^{14})~\Msun$ as a function of projection depth, $\Delta D_p$. The lensing ratios in the ``two-halo" regime reverses trends from low to high projection depth as a result of alignment of clusters with the large scale structure. The profiles with $\Delta D_p = 100~h^{-1}$~Mpc are deemed convergent due to their similarity with the $\Delta D_p = 200~h^{-1}$~Mpc profiles.}
    \label{fig:DeltaSigma_all_vardz}
\end{figure}

The shapes of the profiles can be roughly divided into the ``one-halo" regime ($R \lesssim R_{200m} $) and the ``two-halo" regime ($R \gtrsim R_{200m} $) (Fig. \ref{fig:DeltaSigma_all_vardz}). In the one-halo regime, halos with their major axes oriented towards the line of sight are boosted in their surface density relative to the mean, a result well explained by the triaxial halo model \citep{Oguri05, Corless08}. The transition between the one- and two-halo regimes produces a neck in the surface density, where the halo--matter correlation from neither regime dominates. In the two-halo regime, the trends of the lensing ratios in different orientation become inverted with respect to unity when increasing the projection depth from $\Delta D_p=10h^{-1}$Mpc to $\Delta D_p=200h^{-1}$Mpc. At $\Delta D_p=10h^{-1}$Mpc, the ratio of excess surface densities in the two-halo regime of high $\cos{(i)}$ halos drop below the mean, which may be explained by an under-dense region surrounding the plane perpendicular to the major axes of the halos. As one moves towards larger projection depths, halos with higher $\cos{(i)}$ exhibit boosted $\Delta\Sigma$ profiles in the two-halo regime relative to the mean as a result of the alignment of halos with their underlying large scale structure, i.e., the large projection depth captures much of the mass in the large-scale filaments for halos with $\cos(i)\sim 1$ (\cite{Hahn07b}, \cite{Forero-Romero14}). Because of the similarity of excess surface density profiles for $\Delta D_p = 100~h^{-1}$Mpc and $\Delta D_p=200h^{-1}$Mpc, we deem the projection length $\Delta D_p = 100~h^{-1}$Mpc as convergent. The excess surface density profiles in the one- and two-halo regimes and their dependence on projection depth agree well with \cite{Osato18}, who built profiles for a simulation of similar projections depths and with comparable mass resolution. 

We model the log ratio of excess surface density, $F(R, \mu)$, in a $\mu \equiv \cos(i)$ bin relative to the mean with six free parameters given by the product of a multipole expansion over $\cos(i)$ and a Cauchy function:
\begin{align}
F(R, \mu) &= A(\mu) f(R) \nonumber \\
A(\mu) &= A_0 + A_1 \mu + A_2 \mu^2 + A_3 \mu^3 \nonumber   \\
f(x \equiv \ln(R)) &= 1 - \frac{1}{(x-x_0)^2 + \gamma}.
\label{eqn:DS_template}
\end{align}

\begin{figure*}
	\centering
	\includegraphics[width=1\textwidth]{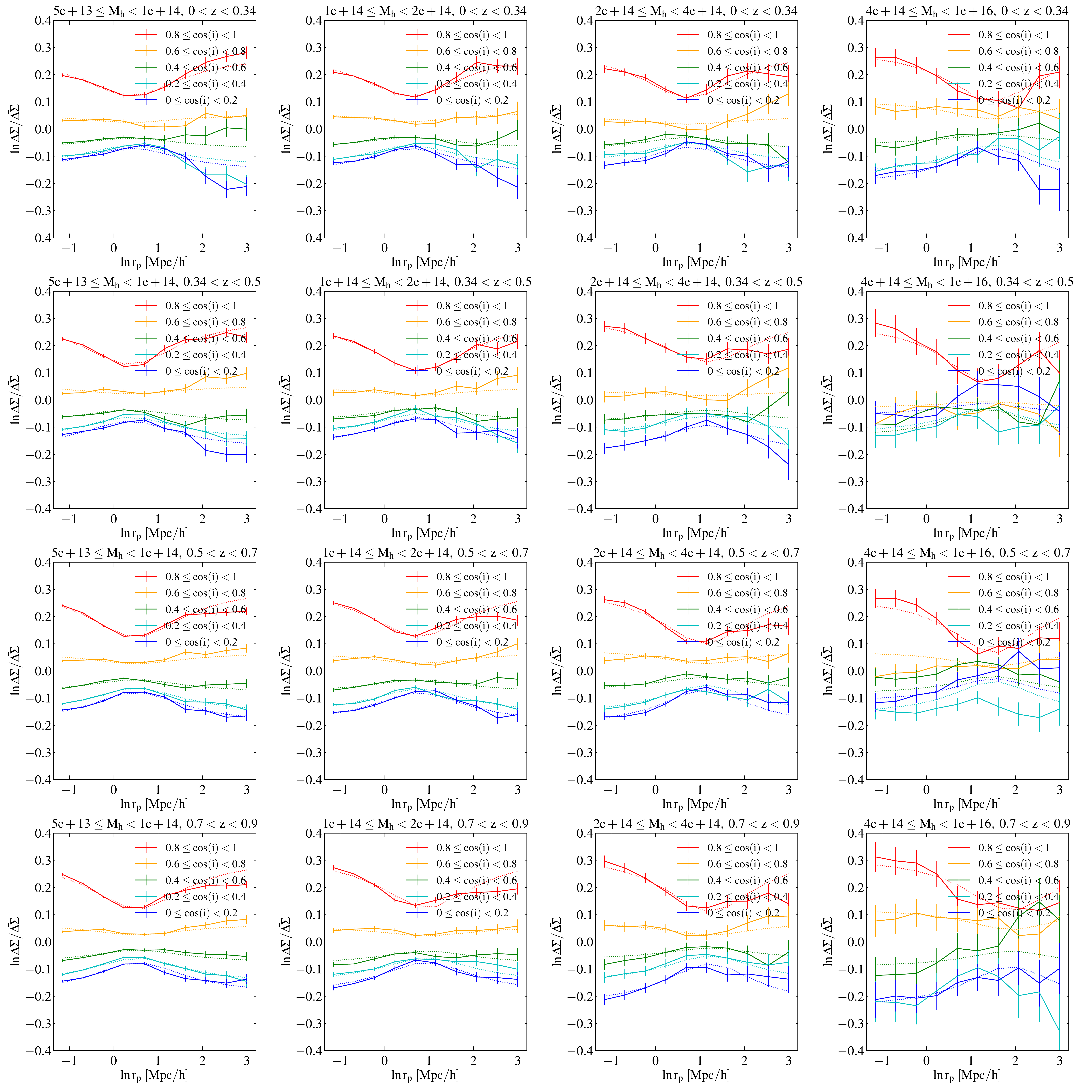}
    \caption{Stacked $\Delta\Sigma$ profiles in different orientations bins (solid lines) vs. Cauchy function fits (dashed lines) to the profiles governed by Equation \ref{eqn:DS_template} and with best fit parameters listed in Table \ref{tab:Cauchy_bestfit}. Error bars are the $1-\sigma$ deviations in measurements in a given orientation and radial bin. }
    \label{fig:fit_cauchy_sqr_errorbar}
\end{figure*}

The bottleneck shape of the $\Delta\Sigma$ profiles binned by $\cos(i)$ is well captured by the Cauchy function in most of the mass and redshift bins, with best-fit parameters and $p$-values listed in Table \ref{tab:Cauchy_bestfit} and plotted in Figure \ref{fig:fit_cauchy_sqr_errorbar}. The parameters show no clear sign of monotonic evolution with mass or redshift that may hint at underlying physics, but they do differ in value from bin to bin, so for greater accuracy the templates are divided into different bins when estimating the stacked mass bias due to triaxiality as will be shown in subsection \ref{subsec:mass_bias}. The best-fit parameters are determined using a Nelder-Mead minimization method; with 10 log-spaced bins in each $\cos{(i)}$ binned $\Delta\Sigma$ profile and 5 $\cos{(i)}$ bins, the templates are fitted with 6 free parameters, totalling $5\times10-6=44$ degrees of freedom; the $\chi^2$ and $p$-value are calculated for each fit. Of the 16 fits, 8 have left- or right-handed $p$-values within 0.01, and 11 within 0.001. The over-fitted templates occur in high-mass or high-redshift bins, which suffer larger errors from the dearth of dark matter particle samples in each bin, and the under-fitted ones result from a mismatch in the ``two-halo regime" that exhibits more poorly constrained trends from bin to bin and the behavior of which is less well understood. Qualitatively, the fits preserve the basic underlying shape of the excess surface density ratios, as shown in Fig. \ref{fig:fit_cauchy_sqr_errorbar}. 

The templates provided could be used as correction terms for Stage III and IV weak lensing cluster surveys such as in the comsoSIS pipeline (\cite{Zuntz14}) for DES-Y3.

\subsection{Modeling the effects of halo orientation on richness-binned excess surface density after redMaPPer selection}
\label{subsec:delta_sigma_bias_redM_selected}
\begin{table*}
\centering
\caption{Best fit parameters for Equation \ref{eqn:DS_template} across different mass and redshift bins.}
\label{tab:DeltaSigma_template_all}
\begin{tabular}{|c|c|c|c|c|c|c|c|c|c|c|c|c|c|}
\hline
$z_{\rm min}$ & $z_{\rm max}$ & $M_{\rm min} (M_\odot)$ & $M_{\rm max} (M_\odot)$ & $A_0$ & $A_1$ & $A_2$ & $A_3$ & $x_0$ & $\gamma$ & $\chi^2$ & \begin{tabular}{@{}c@{}}Left-tail \\ p-value\end{tabular}  & \begin{tabular}{@{}c@{}}Right-tail \\ p-value\end{tabular}\\
\hline
0.00 & 0.34 & $5\times10^{13}$ & $1\times10^{14}$ & -0.157 & -0.001 & 0.091 & 0.485 & 1.346 & 0.378 & 85.204 & 0.9998 & 0.0002\\
0.00 & 0.34 & $1\times10^{14}$ & $2\times10^{14}$ & -0.168 & -0.0107 & 0.222 & 0.375 & 1.325 & 0.592 & 25.580 & 0.012 & 0.988 \\
0.00 & 0.34 & $2\times10^{14}$ & $4\times10^{14}$ & -0.197 & 0.373 & -0.818 & 1.112 & 1.289 & 0.757 & 34.009 & 0.139 & 0.861 \\
0.00 & 0.34 & $4\times10^{14}$ & $1\times10^{16}$ & -0.190 & -0.270 & 1.307 & -0.457 & 1.245 & 1.504 & 18.551 & 0.9997 & 0.0003\\
0.34 & 0.50 & $5\times10^{13}$ & $1\times10^{14}$ & -0.204 & 0.264 & -0.489 & 0.909 & 1.320 & 0.403 & 65.605 & 0.981 & 0.020 \\
0.34 & 0.50 & $1\times10^{14}$ & $2\times10^{14}$ & -0.190 & 0.238 & -0.472 & 0.888 & 1.261 & 0.782 & 23.623 & 0.995 & 0.005\\
0.34 & 0.50 & $2\times10^{14}$ & $4\times10^{14}$ & -0.281 & 0.952 & -2.056 & 1.913 & 1.342 & 1.141 & 25.567 & 0.988 & 0.012 \\
0.34 & 0.50 & $4\times10^{14}$ & $1\times10^{16}$ & -0.021 & -0.268 & -0.681 & 1.504 & 1.146 & 1.344 & 28.903 & 0.962 & 0.038  \\
0.50 & 0.70 & $5\times10^{13}$ & $1\times10^{14}$ & -0.212 & 0.190 & -0.174 & 0.669 & 1.292 & 0.523 & 91.768 & 1.000 & 0.000\\
0.50 & 0.70 & $1\times10^{14}$ & $2\times10^{14}$ & -0.203 & 0.103 & 0.017 & 0.547 & 1.307 & 0.784 & 66.490 & 0.9841 & 0.016 \\
0.50 & 0.70 & $2\times10^{14}$ & $4\times10^{14}$ & -0.214 & 0.095 & 0.213 & 0.350 & 1.228 & 1.126 & 48.257 & 0.305 & 0.695 \\
0.50 & 0.70 & $4\times10^{14}$ & $1\times10^{16}$ & -0.036 & -0.996 & 2.188 & -0.780 & 1.148 & 1.514 & 88.200 & 0.9999 & 0.0001\\
0.70 & 0.90 & $5\times10^{13}$ & $1\times10^{14}$ & -0.208 & 0.209 & -0.263 & 0.738 & 1.290 & 0.564 & 99.745 & 1.000 & 0.000\\
0.70 & 0.90 & $1\times10^{14}$ & $2\times10^{14}$ & -0.213 & 0.305 & -0.612 & 1.030 & 1.29 & 0.931 & 71.863 & 0.995 & 0.005\\
0.70 & 0.90 & $2\times10^{14}$ & $4\times10^{14}$ & -0.287 & 0.655 & -0.975 & 1.105 & 1.243 & 1.226 & 33.589 & 0.873 & 0.127 \\
0.70 & 0.90 & $4\times10^{14}$ & $1\times10^{16}$ & -0.158 & -1.184 & 3.551 & -1.839 & 1.260 & 1.904 & 21.298 & 0.9985 & 0.0015\\
\hline
\end{tabular}
\label{tab:Cauchy_bestfit}
\end{table*}

Stacking refers to the process of building averaged excess surface density profiles of halos in different richness bins. This subsection describes the process of stacking used by the DES survey to calibrate the richness--mass relation and presents the effect of triaxiality on the stacked surface density. 

The shapes of source galaxies behind a cluster along the line of sight will have small tangential distortions due to gravitational lensing. While individual distortions are small, this tangential shear can be measured at high signal to noise as a function of projected radial separation $R$ in the stacked images of source galaxies around clusters binned, e.g., in richness and redshift. In the weak lensing regime, the tangential shear is related to the source-galaxy ellipticity by
\beqa
\gamma_t \approx e^{T} + \rm noise,
\eeqa
where $e^{T}$ is the source ellipticity rotated to the tangential frame, and the noise is due to intrinsic ellipticities of the source galaxies (shape noise) and measurement uncertainty. The tangential shear, $\gamma_t$, as directly measured by observations can be converted to $\Delta\Sigma(R)$ through Equation \ref{eqn:DeltaSigma}. This paper directly measures $\Delta\Sigma(R)$ by computing the 2D dark-matter density along a cylinder of given projection depth centered around the cluster.

The model excess surface density is obtained by integrating the halo--matter correlation $\xi_{hm}(r)$ along the line of sight as in equation \ref{eqn:halo_matter_corr}, and subtracting that from the mean surface density inside the projected radius as in equation \ref{eqn:DeltaSigma}. Typically, the halo--matter correlation in the ``one-halo'' regime is modeled as a spherical Navarro-Frenk-White (NFW) \citep{NFW96} profile $\rho_{\rm NFW}(r|M)$,
\beqa
\xi_{\rm 1h} (r|M) = \frac{\rho_{\rm NFW}(r|M)}{\rho_{m0}} - 1,
\label{eqn:xi_NFW}
\eeqa
and the ``two-halo" term as a linear
matter correlation \citep{Hayashi08} scaled by the halo bias, (e.g. \cite{Tinker10}):
\beqa
\xi_{\rm 2h} (r|M) = b^2(M) \xi_{\rm lin}(r).
\label{eqn:xi_2h}
\eeqa

At the transition between the two regimes, DES Y1 follows \cite{Zu14} in setting the halo--matter correlation to the maximum value of the two terms, i.e.,
\beqa
\xi_{hm}(r|M) = \rm max~\{\xi_{\rm 1h}(r|M),~\xi_{\rm 2h}(r|M) \}
\label{eqn:xi_max}
\eeqa

In our analysis we reproduce the surface density templates from the procedures in the DES Y1 analysis using publicly available code---the 
linear power spectrum computed from CLASS (\cite{Lesgourgues11, Blas11}) and the excess surface density computed from the \textit{cluster\_toolkit} module\footnote{Code written by Tom McClintock. Source: http://cluster-toolkit.readthedocs.io/en/latest/index.html},  which uses the spherical NFW profile for the ``one-halo'' term and refers to \cite{Tinker10} for the halo bias---to generate isotropic profiles, which we denote $\overline{\DS}(R)$, calculated by integrating through Equation \ref{eqn:halo_matter_corr} $\xi_{hm}$ in the form of Equation \ref{eqn:xi_max}. In the ``one-halo" regime we parametrize the NFW profile with a nominal concentration of $c=5$.

We investigate the difference in the stacked profile between the isotropic $\overline{\DS}(R)$ and $\overline{\Delta\Sigma}(R, M, \mu)$, the stacked profile as a function of orientation dependence. 

The orientation dependence has two components---one is the scaling of individual lensing profiles by $\exp(F(R,\mu))$ as described in Section \ref{subsec:delta_sigma_profile}, and the other the effect of richness-mass, $P(\lambda|M, \mu)$, as modeled in Section \ref{sec:richness--mass}, on the mass distribution of redMaPPer-selected clusters. 
The second component, $P(\lambda|M, \mu)$, biases the mass distribution of clusters in a richness bin $\widetilde{P}(M)$ through the form
\begin{align}
\widetilde{P}(M) &= \int d\mu \int^{\lambda_2}_{\lambda_1} d\lambda P(M, \lambda, \mu)  \nonumber \\
&= \int d\mu \int^{\lambda_2}_{\lambda_1} d\lambda P(\lambda | M, \mu) P(\mu|M) P(M) \nonumber \\
&\text{and safely assuming that $P(\mu|M)$ is constant,} \nonumber \\
&\propto \int d\mu \int^{\lambda_2}_{\lambda_1} d\lambda P(\lambda | M, \mu)  P(M), 
\end{align}
where $P(M)$ is the mass function of redMaPPer-selected clusters.

The conditional probability of richness, $P(\lambda|M,\mu)$, is log-normally distributed around a mean richness governed by Equation \ref{eqn:richness_mass_mean}, and the standard deviation is given by equation \ref{eqn:richness_mass_scatter}. The equations are fit to the one-parameter model in which only $\log(A)$, the intercept of the $\log(\lambda)$-$\log(M)$ relation, is allowed to vary with orientation. We use a cubic spline to interpolate $\log(A)$ for $\mu \in [0,1)$. The halo-mass function of redMaPPer-selected clusters, $P(M)$, is constructed from a discrete histogram with 30 log-spaced mass bins in the mass range of the clusters. 

Taking into account the two components for orientation dependence, the stacked surface density in a richness bin becomes
\begin{align}
&\overline{\Delta\Sigma}(R, M, \mu)  ~\text{for}~\lambda \in [\lambda_1, \lambda_2)  \nonumber \\
&= \int dM~\Delta\Sigma(R, M, \mu) \widetilde{P}(M)  \nonumber \\
 & = \int d\mu \int dM \int^{\lambda_2}_{\lambda_1} d\lambda ~\Delta\Sigma(R, M, \mu) P(\lambda | M, \mu) P(\mu|M) P(M)  \nonumber \\
 &\propto \int d\mu \int dM \int^{\lambda_2}_{\lambda_1} d\lambda ~\Delta\Sigma(R, M, \mu) P(\lambda | M, \mu) P(M) 
\label{eqn:DS_stacked}
\end{align}

The excess surface densities are computed for $\avg{\DS(M,R,\mu)}$ using equation \ref{eqn:DS_stacked} and $\overline{\DS}(M,R)$ using equations \ref{eqn:xi_NFW}--\ref{eqn:xi_max}. We define the fractional difference with the shorthand notation
\begin{equation}
    \delta\langle\Delta\Sigma\rangle = \frac{\overline{\Delta\Sigma}(R,\mu) - \overline{\Delta\Sigma}(R)}{\overline{\Delta\Sigma}(R)}.
\end{equation}

\subsection{Mass bias estimation of stacked clusters}
\label{subsec:mass_bias}
We are interested in estimating the effect of triaxiality  on the mean weak lensing mass in clusters stacked in richness bins. The weak lensing mass is an observed quantity in weak lensing surveys derived by fitting the observed lensing profile to an analytic profile in a procedure akin to that in Section \ref{subsec:delta_sigma_bias_redM_selected} and is used to constrain the mass-richness relation. We estimate the bias due to triaxiality on the weak lensing mass for stacked clusters by propagating the error on the lensing observable onto the mass model parameter using a Fisher matrix approximation. 

In the most generic sense, the Fisher matrix $F_{ij}$ in a given radial bin is defined as:
\begin{equation}
F_{ij}(R) = \frac{\partial \langle\Delta\Sigma\rangle(R)}{\partial p_i} {\rm Cov(\langle\DS\rangle(R))}^{-1} \frac{\partial \langle\Delta\Sigma\rangle(R)}{\partial p_j},
\end{equation}
where the partial derivatives are of surface density profiles with respect to model parameters $p_i$ of cluster mass $M$ and concentration $c$, and the covariance matrix is that of surface density as a function of radius.

The mass-bias for stacked clusters due to triaxiality is given by the expression
\begin{equation}
    \delta M_{\rm binned} = 
    \sum_{j} (F^{-1})_{ij}
    \left[(\delta\langle\Delta\Sigma\rangle) {\rm Cov(\langle\DS\rangle)^{-1}} \frac{\partial \Delta\Sigma}{\partial p_j}\right],
\end{equation}
estimated by inserting the fractional difference of stacked profiles, $\delta(\Delta \Sigma)$, into the bracketed expression and marginalizing over the concentration parameter. The total bias is then the weighted sum of all mass and redshift bins marginalized over concentration and radius:
\begin{equation}
    \delta M_{\rm total} = 
    \sum_{M,z} P(M,z | \lambda) \left[ \sum_{j,R} (F^{-1})_{ij}(R)
     \left(\delta\langle\Delta\Sigma\rangle {\rm Cov(\langle\DS\rangle)^{-1}} \frac{\partial \Delta\Sigma}{\partial p_j}\right)\right].
\label{eqn:mass_bias_tot}
\end{equation}

The $\langle\Delta\Sigma\rangle$ profiles are binned in richness intervals of $\lambda\in[20,~30),~ [30,~50),~\rm and~[50,~\infty)$, and are further divided into the same mass and redshift bins when computing individual $\Delta\Sigma (R)$ templates as described in Section \ref{subsec:delta_sigma_profile}. We make the simplifying assumption that the partial derivative of the bin-averaged surface density profile is well approximated by that for a numerical model for an individual halo, with $M$ taken at the midpoint of the mass bin, and $c$ derived from redshift and mass using the relation
\begin{equation}
    c_{200b} = \frac{c_0}{1+z}\Big(\frac{M}{M_0}\Big)^{-\beta},
    \label{eqn:concentration_M08}
\end{equation}
with functional form and best fit parameters of $c_0 = 4.6$ at $z=0.22$ and $\beta = 0.13$ from \cite{Mandelbaum08}, calculated at the midpoint value of said mass bin. The 
approximation of $\langle \Delta\Sigma(R) \rangle$ profiles is computed using \textit{cluster\_toolkit} for the Buzzard cosmological parameters. 

The covariance matrix for cluster weak lensing is taken from \cite{Wu19}, who calculated the matrices from a combination of analytic calculations and high-resolution N-body simulations for radii between 0.1 and 100 $h^{-1}~\rm Mpc$, discretized at 15 equally log-spaced bins. The covariance comes from a combination of shape noise, large scale structure and intrinsic noise. Modeled on a DES-like simulation with a galaxy density of $n_s \sim 10/$arcmin, the covariance is dominated by shape noise at projected radii $\lessapprox~ 5h^{-1}$~Mpc. The covariance matrices are binned by mass in bins of  $[10^{14},~2\times10^{14}),~[2\times10^{14},~4\times10^{14})~\rm and$ $[4\times10^{14},~\infty)~h^{-1}M_{\odot}$, and in lens/source redshift slices of $\{z_l = 0.3, z_s = 0.75\}$, $\{z_l = 0.5, z_s = 1.25\}$ and $\{z_l = 0.7, z_s = 1.75\}$, with $z_l$ denoting the lens redshift and $z_s$ the source redshift.

To address the different binning schemes used in the lensing covariance and stacked lensing profiles, we choose to evaluate the covariance at the central redshift slice of $\{z_l = 0.5, z_s = 1.25\}$, since the redshift dependence of the lensing covariance is weak. Because the covariance matrix is not applicable for masses below $10^{14}~h^{-1} M_{\odot}$, we ignore $\langle \Delta\Sigma(M,z) \rangle$ in the modeling for Equation \ref{eqn:mass_bias_tot} for the lowest mass bin of $[5\times10^{13},~1\times10^{14})~h^{-1}M_{\odot}$. Making this mass cut removes $35\%$ of the redMaPPer clusters in total. 

\begin{figure}
	\centering
	\includegraphics[width=0.4\textwidth]{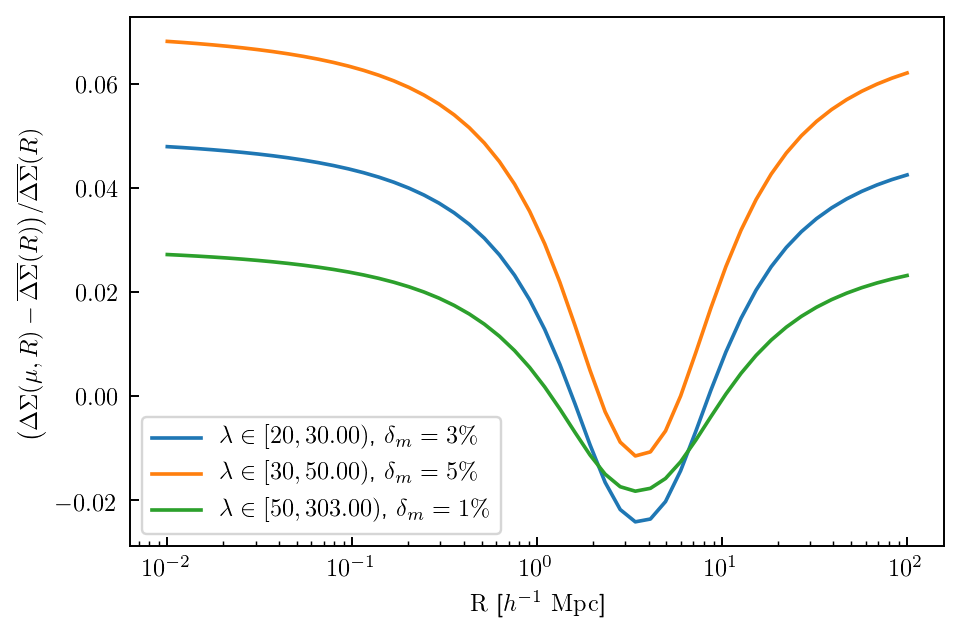}
    \caption{Fractional difference in lensing profiles $\delta\langle\Delta\Sigma\rangle$ for redMaPPer-selected clusters stacked in bins of richness. The total mass bias for each richness bin is measured by marginalizing $\delta\langle\Delta\Sigma\rangle$ as shown in plot through Equation \ref{eqn:mass_bias_tot} through propagating the errors of the lensing profile onto the mass model parameter using a Fisher forecast. }
    \label{fig:stacked_tot_mass_bias}
\end{figure}

Using the covariance matrix from \cite{Wu19} and the mass-concentration relation of \cite{Mandelbaum08}, we calculate the total mass bias through the propagation of bias from the lensing signal onto the mass model parameter through a Fisher matrix forecast. As shown in Figure \ref{fig:stacked_tot_mass_bias}, the mass is biased high at $1-5 \%$, consistent with findings from \cite{McClintock18} and \cite{Dietrich14} and is highest at mid-richness ranges. 

Our results are consistent with the recently released DES Y1 cluster cosmology paper \citep{DESY1_Cluster}, which tested for systematics by controlling for variables that may introduce bias. The lensing profiles of two samples were compared---one selected by richness bins with its mass distribution left free to vary, and the second tracing the mass distribution of the richness-selected sample with its richness free to vary. The ratio of these profiles is an estimate of the total systematic bias due to redMaPPer selection in a given richness bin and radial range. The effects of triaxiality and projection effects can be teased out by re-sampling their proxies $\cos{(i)}$ and $\sigma{(z)}$ in the richness-selected sample to match the mass-selected sample. The paper showed that triaxiality and projection effects were capable of adjusting the ratios of the two lensing profiles to unity within errors for richness $\lambda > 30$ but failed to resolve the tension at $\lambda =$~20-30. The maximal impact of triaxiality at mid-to-high richness ranges supports the finding that as triaxiality bias weakens at low-richness, some other unaccounted-for systematic must be in play.

\section{Conclusion}
The main findings of this work are as follows:
\begin{enumerate}
    \item We find that the prolateness distribution of redMaPPer-selected halos is consistent with the prolateness distribution of halos overall.
    
    \item We find that the log-richness amplitude $\ln(A)$ of redMaPPer clusters for a given mass is boosted from the lowest to highest orientation bin with a significance of 14$\sigma$.
    
    \item We find a null correlation between the bias in richness due to triaxiality and those for two other leading systematics in DES Y1 cluster cosmology---miscentering and projection---and offer explanations or follow-up studies for this result. The  null correlation with projection effects is was verified using both the Buzzard and C19 projection mock, catalogs with different galaxy-halo connection models.
    
    \item We confirm the bottleneck shape in the transition between one- and two-halo regimes for halo lensing profiles first discovered by \cite{Osato18} and fit it to redshift- and mass-dependent templates. 
    
    \item We quantify through items (ii) and (iv) the DES observable of richness-stacked redMaPPer cluster lensing profiles to predict a positive mass bias of $1-5\%$ due to triaxiality. 
    
    \item We find that the mean $P(\cos{i})$ and the mass bias are both richness dependent and largest at  mid-to-high richness, in accordance with the DES Y1 result that triaxiality does not fully resolve the tension in weak lensing mass at low richness. 
\end{enumerate}

Our findings are based on redMaPPer catalogs constructed using galaxies in the Buzzard simulations. The realistic red-sequence galaxy model in the Buzzard simulations allows us to run the redMaPPer algorithm in the same way as it was run on DES-Y1 data and hence enables us to quantify various selection effects introduced by the cluster finder. While this analysis provides evidence of redMaPPer selection effects and quantifies the relations between different systematics, we must acknowledge that there is one important caveat in this approach: the performance of the redMaPPer cluster finder depends on how galaxies are populated in the simulations, which might not precisely match the real universe. Since this analysis is only done on one specific simulation, the result in this paper can serve as guidance for constructing a flexible enough model used in the analysis of real data.

These findings shed light on the impact of triaxiality on cluster selection, both their physical quantities and observed signals. Specifically, items (ii) and (iv) may be used as templates for current and near future weak lensing surveys as correction terms for this systematic. One important future work is to perform this analysis on different mock galaxy catalogs with different assumptions about the relations between galaxies and dark matter. Such an analysis will be essential to addressing the dependence of cluster finder performance on galaxy population models.

\section*{Acknowledgements}
The authors acknowledge Tae-hyeon Shin, Andrew Hearin, Chihway Chang and William Wester, among many others for their input to this work.

Research was supported by the U.S. Department of Energy (DOE) Office of Science Distinguished Scientist Fellow Program.

We would like to thank Stanford University and the Stanford Research Computing Center for providing computational resources and support that contributed to these research results. This research used resources of the National Energy Research Scientific Computing Center (NERSC), a U.S. Department of Energy Office of Science User Facility operated under Contract No. DE-AC02- 05CH11231.

Funding for the DES Projects has been provided by
the U.S. Department of Energy, the U.S. National Science Foundation, the Ministry of Science and Education
of Spain, the Science and Technology Facilities Council of the United Kingdom, the Higher Education Funding Council for England, the National Center for Supercomputing Applications at the University of Illinois at
Urbana-Champaign, the Kavli Institute of Cosmological
Physics at the University of Chicago, the Center for Cosmology and Astro-Particle Physics at the Ohio State University, the Mitchell Institute for Fundamental Physics
and Astronomy at Texas A\&M University, Financiadora
de Estudos e Projetos, Funda¸c˜ao Carlos Chagas Filho
de Amparo `a Pesquisa do Estado do Rio de Janeiro,
24
Conselho Nacional de Desenvolvimento Cient´ıfico e Tecnol´ogico and the Minist´erio da Ciˆencia, Tecnologia e Inova¸c˜ao, the Deutsche Forschungsgemeinschaft and the
Collaborating Institutions in the Dark Energy Survey.

The Collaborating Institutions are Argonne National
Laboratory, the University of California at Santa Cruz,
the University of Cambridge, Centro de Investigaciones
Energ´eticas, Medioambientales y Tecnol´ogicas-Madrid,
the University of Chicago, University College London,
the DES-Brazil Consortium, the University of Edinburgh, the Eidgen¨ossische Technische Hochschule (ETH)
Z¨urich, Fermi National Accelerator Laboratory, the University of Illinois at Urbana-Champaign, the Institut de
Ci`encies de l’Espai (IEEC/CSIC), the Institut de F´ısica
d’Altes Energies, Lawrence Berkeley National Laboratory, the Ludwig-Maximilians Universit¨at M¨unchen and
the associated Excellence Cluster Universe, the University of Michigan, the National Optical Astronomy Observatory, the University of Nottingham, The Ohio State
University, the University of Pennsylvania, the University of Portsmouth, SLAC National Accelerator Laboratory, Stanford University, the University of Sussex, Texas
A\&M University, and the OzDES Membership Consortium.

Based in part on observations at Cerro Tololo InterAmerican Observatory, National Optical Astronomy Observatory, which is operated by the Association of Universities for Research in Astronomy (AURA) under a cooperative agreement with the National Science Foundation.
The DES data management system is supported by
the National Science Foundation under Grant Numbers AST-1138766 and AST-1536171. The DES participants from Spanish institutions are partially supported
by MINECO under grants AYA2015-71825, ESP2015-
66861, FPA2015-68048, SEV-2016-0588, SEV-2016-0597,
and MDM-2015-0509, some of which include ERDF funds
from the European Union. IFAE is partially funded by
the CERCA program of the Generalitat de Catalunya.
Research leading to these results has received funding
from the European Research Council under the European Union’s Seventh Framework Program (FP7/2007-
2013) including ERC grant agreements 240672, 291329,
and 306478. We acknowledge support from the Australian Research Council Centre of Excellence for Allsky Astrophysics (CAASTRO), through project number
CE110001020.

This manuscript has been authored by Fermi Research
Alliance, LLC under Contract No. DE-AC02-07CH11359
with the U.S. Department of Energy, Office of Science,
Office of High Energy Physics. The United States Government retains and the publisher, by accepting the article for publication, acknowledges that the United States
Government retains a non-exclusive, paid-up, irrevocable, world-wide license to publish or reproduce the published form of this manuscript, or allow others to do so,
for United States Government purposes.




\newpage
\bibliographystyle{mnras}
\bibliography{master_refs} 

\bsp	
\label{lastpage}
\end{document}